\documentclass{article}

\PassOptionsToPackage{numbers, compress}{natbib}

    \usepackage[preprint]{neurips_2021}

\usepackage{algorithm, algpseudocode}
\usepackage[utf8]{inputenc} 
\usepackage[T1]{fontenc}    
\usepackage{hyperref}       
\usepackage{url}            
\usepackage{booktabs}       
\usepackage{amsfonts}       
\usepackage{nicefrac}       
\usepackage{microtype}      
\usepackage{xcolor}         
\usepackage{amsmath, amsopn, amsthm, epsfig}
\usepackage{subfigure}
\usepackage{graphicx, epstopdf}

\title{It\^oTTS and It\^oWave: Linear Stochastic Differential Equation Is All You Need For Audio Generation}

\author{%
  Shoule Wu\\
  Yangzhou University\\
 \texttt{wu.shoule@protonmail.com} \\
   \And
   Ziqiang Shi\\
   Fujitsu R\&D Center \\
   \texttt{shiziqiang@fujitsu.com} \\
}

\begin{document}

\maketitle

\begin{abstract}
  In this paper, we propose to unify the two aspects of voice synthesis, namely text-to-speech (TTS) and vocoder, into one framework based on a pair of forward and 
  reverse-time linear stochastic differential equations (SDE). The solutions of this SDE pair are two stochastic processes, one of which turns the distribution of 
  mel spectrogram (or wave), that we want to generate, into a simple and tractable distribution. The other is the generation procedure that turns this tractable 
  simple signal into the target mel spectrogram (or wave). The model that generates mel spectrogram is called It\^oTTS, and the model that generates 
  wave is called It\^oWave. It\^oTTS and It\^oWave use the Wiener process as a driver to gradually subtract the excess 
  signal from the noise signal to generate realistic corresponding meaningful mel spectrogram and audio respectively, under the conditional inputs of original 
  text or mel spectrogram. The results of the experiment show that the mean opinion scores (MOS) of   It\^oTTS and It\^oWave can exceed 
  the current state-of-the-art methods, and reached 3.925$\pm$0.160 and 4.35$\pm$0.115 respectively.
  The generated audio samples are available at https://shiziqiang.github.io/ito\_audio/.

  All authors contribute equally to this work.
\end{abstract}

\section{Introduction}
\label{sec:intro}

In recent years, the generation technology, especially the voice generation technology has made great progress. 
Voice generation technology consists of two important and relatively independent modules, one is text-to-speech (TTS) and the other is vocoder. 
Among them, TTS transforms text into voice features, such as mel spectrogram, and vocoder transforms voice features into waveforms.
Most researchers study TTS and vocoder separately, and design for different characteristics of different tasks. 

Whether it is a TTS or a vocoder, the model is roughly categorized as autoregressive (AR) or non-autoregressive (non-AR), where the AR model generates the signal 
frame by frame, and the generation of the current signal frame depends on the previously generated signal.  Non-AR models generate the signal in parallel, and the 
current signal frame does not depend on the previous signal. Generally speaking, the voice quality generated by the AR model is 
higher than the non-AR model, but the amount of computation is also larger, and the generation speed is slow. While for the non-AR generation model, the generation 
speed is faster, 
but the generated voice quality is slightly worse. To name a few, for example, in the field of TTS, AR-type models are generally implemented through the framework of 
encoder and decoder, and the duration of text is predicted 
by an attention module, such as Tacotron~\citep{wang2017tacotron} and Tacotron 2~\citep{shen2018natural}, DeepVoice 2~\citep{gibiansky2017deep} and 3~\citep{ping2018deep}. 
In the last two years, most of the work has focused on the study of non-AR TTS models, such as flow based models~\citep{valle2020flowtron,miao2020flow,kim2020glow}, 
variational auto-encoder (VAE) based models~\citep{liu2021vara}, 
generative adversarial network (GAN) based models~\citep{binkowski2019high}, and Fastspeech and Fastspeech 2~\citep{ren2019fastspeech,ren2020fastspeech} etc.

The situation in the vocoder field is similar. WaveNet~\citep{oord2016wavenet} is the earliest AR model, using sampling points as the unit and achieves a sound quality 
that matches the naturalness of human speech. In addition, other recent AR models, including sampleRNN~\citep{mehri2016samplernn} and LPCNet~\citep{valin2019lpcnet}
have further improved the sound quality. 
However, due to the large amount of 
computation and the slow generation speed, researchers currently mainly focus on developing non-AR wave generation models, such as Parallel WaveNet~\citep{oord2018parallel}, 
ClariNet~\citep{ping2018clarinet}, GanSynth~\citep{engel2019gansynth}, FloWaveNet~\citep{kim2018flowavenet}, MelGan~\citep{kumar2019melgan}, WaveGlow~\citep{prenger2019waveglow}, 
Parallel WaveGan~\citep{yamamoto2020parallel}, and so on.

In this paper, TTS and vocoder are modeled with the same new framework based on linear It\^o stochastic differential equations (SDE) and score matching modeling. 
We call them It\^oTTS and It\^oWave respectively. The linear It\^o SDE, driven by the Wiener process, can slowly 
turn the mel spectrogram and wave data distributions, that need to be generated in TTS and vocoder, into data distributions that are easy to manipulate, such as white noise. 
This transformation process is the stochastic process solution of the linear It\^o SDE. Therefore, the corresponding  reverse-time linear 
It\^o SDE can generate the mel spectrogram and wave data distribution required by TTS 
and vocoder,  from this easy data distribution, such as white noise. It can be seen that the reverse-time linear It\^o SDE is crucial for the generation, 
and~\citet{anderson1982reverse} shows the 
explicit form of this reverse-time linear SDE, and the formula shows that it depends on the gradient of the log value of the probability density function 
of the stochastic process solution of the forward-time equation. This gradient value is also called the stein score~\citep{hyvarinen2005estimation}. 
It\^oTTS and It\^oWave predict the stein score corresponding to the mel spectrogram or wave by trained neural networks. 
After obtaining this score, It\^oTTS and It\^oWave can achieve 
the goal of generating mel spectrogram and wave through reverse-time linear It\^o SDE or Langevin dynamic sampling.

Our contribution is as follows:
\begin{itemize}
  \item We are among the first\footnotemark[1] to proposed a TTS and vocoder model based on linear It\^o SDE, called It\^oTTS and It\^oWave, 
  which reached state-of-the-art performance. These 
models are easier to train than the GAN-based models and do not require a reversible network like the flow models.
\item We explicitly unify TTS and vocoder under a more flexible framework, which can construct different 
TTS and vocoder models by selecting different  drift and diffusion coefficients of the linear SDE.
\item For It\^oTTS and It\^oWave, we propose two network structures, which are suitable for estimating the gradient of log value of 
the density function of the 
mel spectrogram and wave data distributions.
\item It is believed that It\^oTTS and It\^oWave provide many knowhows, serves as a good and new high starting baseline, 
and there will be more linear SDE-based sound generation work in the future.
\end{itemize}

\footnotetext[1]{In May 2021, we found that another  
group has also developed a TTS algorithm call Grad-TTS~\citep{popov2021gradtts} based on SDE, but our research and development are 
completely independent of theirs, and they didn't present vocoder method. Our advantage compared with theirs, is that both TTS and vocoder in our method are implemented with 
the same framework based on linear SDE, and our score predict networks are different from theirs.  
It\^oTTS’s score prediction network uses a dilated residual network for its decoder, which is more flexible and longer
 than Grad-tts’ U-Net in terms of the receptive field, and has a low computational load.}

\section{Related work}
\label{sec:related}

The earliest source of the ideas for It\^oTTS and It\^oWave should be the pioneering change of data estimation problem into the estimation of 
the gradient of log of the data distribution 
density by~\citet{hyvarinen2005estimation}, thus greatly simplifying the original problem. Another source is the pioneering use of a diffusion Markov chain
 by~\citet{sohl2015deep} to diffuse 
the structure of the image data into a simple distribution, and another opposite diffusion Markov chain to generate images in the target distribution from the 
simple distribution. In the past two years, these two primitive ideas have been carried forward. ~\citet{ho2020denoising} has recently generated very 
high-quality large-scale natural 
images with a diffusion Markov chain. DiffWave~\citep{kong2020diffwave} uses a diffusion Markov chain for the vocoder. ~\citet{song2019generative} draws on the idea of 
~\citep{hyvarinen2005estimation} to estimate the log gradient 
of the target data distribution 
density function through a neural network, and then uses the Langevin dynamics to generate large-scale image data in target distribution. 
Wavegrad~\citep{chen2020wavegrad} transplanted the algorithm of~\citet{song2019generative} to vocoders, but 
in fact the final algorithm is exactly the same as Diffwave~\citep{kong2020diffwave}. Immediately afterward, ~\citet{song2020score} further extended the
 Markov chain to the continuous case, 
it became a stochastic 
differential equation. Ingeniously, the equation unified the two methods of~\citet{ho2020denoising}  and~\citet{song2019generative}  under one framework, 
and both became its special cases.

It\^oTTS and It\^oWave successfully found the realization of the linear SDE framework in sound generation and achieved very high-quality sound.

\section{It\^oTTS and  It\^oWave}

\subsection{Audio data distribution transformation based on It\^o SDE}
\label{sec:dist_transormation_sde}

It\^o SDE is a very natural model that can realize the transformation between different data distributions. The general 
It\^o SDE is as follows
\begin{align}
  \left\{
  \begin{array}{lr}
 d\mathbf{X}=\mathbf{f}(\mathbf{X},t)dt+g(t)d\mathbf{W} &  \\
 \mathbf{X}(0)=\mathbf{x}(0) &
\end{array}
\right.
 \label{eq:ito_sde}
\end{align}
for $0\leq t\leq T$, where  $\mathbf{f}(\cdot ,t)$ is the drift coefficient,
 $g(t)$ is the diffusion coefficient,
 $\mathbf{W}$ is the standard Wiener process. The solution of this It\^o SDE is a stochastic process
 $\mathbf{X}(t)$ satisfies the following It\^o integrals
 \begin{align}
 \mathbf{X}(t)=\mathbf{x}(0)+\int_0^t\mathbf{f}(\mathbf{X}(s),s)ds+\int_0^tg(s)d\mathbf{W}(s)
 \label{eq:ito_sde_integral}
\end{align}
all most surely for all $0\leq t\leq T$. Let $p(\mathbf{x}(t))$ be the density of the random variable $\mathbf{X}(t)$. 
This SDE~(\ref{eq:ito_sde}) changes the initial distribution $p(\mathbf{x}(0))$ into another distribution 
$p(\mathbf{x}(T))$
by gradually adding the noise from the Wiener process $\mathbf{W}$.
In this work, $\mathbf{x}(t)\in\mathbb{R}^d$, and $p(\mathbf{x}(0))$ is to denote the data distribution of mel spectrogram or wave in It\^oTTS 
or It\^oWave 
respectively. 
$p(\mathbf{x}(T))$ is an easy tractable distribution (e.g. Gaussian) of the latent representation of the mel spectrogram or wave signal corresponding to the conditional text or 
mel spectrogram.
If this stochastic process $\mathbf{x}(t)$  can be reversed in time, then the target mel spectrogram or waveform of the corresponding text or mel spectrogram can be generated
  from a simple latent distribution. 

Actually the reverse-time diffusion process is the solution  of the following corresponding reverse-time It\^o SDE~\citep{anderson1982reverse}
\begin{align}
  \left\{
  \begin{array}{lr}
 d\mathbf{X}=\left[\mathbf{f}(\mathbf{X},t)-g(t)^2\nabla_\mathbf{x}\log p(\mathbf{x}(t))\right]dt
 +g(t)d\overline{\mathbf{W}}  &  \\
 \mathbf{X}(T)=\mathbf{x}(T) &
\end{array}
\right.
 \label{eq:reverse_ito_sde}
\end{align}
for $0\leq t\leq T$,
where $p(\mathbf{x}(t))$ is  the distribution of $\mathbf{X}(t)$, 
$\overline{\mathbf{W}}$ is the standard Wiener process in reverse-time.
The solution of this reverse-time It\^o SDE~(\ref{eq:reverse_ito_sde}) can be used to generate 
mel spectrogram or wave data from a tractable  latent distribution $p(\mathbf{x}(T))$. 
Therefore, it can be seen from~(\ref{eq:reverse_ito_sde}) that the key to generating mel spectrogram or wave with SDE lies in the calculations of 
$\nabla_\mathbf{x}\log p(\mathbf{x}(t))$ $(0\leq t\leq T)$, which is always called score function~\citep{hyvarinen2005estimation,song2019generative} of the data.

\subsection{Score estimation of audio data distribution}
\label{sec:score_estimation}


In this work, a neural network $\mathfrak{S}_{\theta}$ is used to approximate the score function, where $\theta$ denotes the parameters of the network. 
The input of the network $\mathfrak{S}_{\theta}$  includes time $t$, $\mathbf{x}(t)$, and conditional input text $\mathbf{p}$ or mel spectrograms $\mathbf{m}$ in
 It\^oTTS and It\^oWave 
respectively.
The expected output 
is $\nabla_{\mathbf{x}(t)} \log  p(\mathbf{x}(t))$.  
The objective of score matching is~\citep{hyvarinen2005estimation} (here we only take It\^oWave as an example, 
the situation of It\^oTTS is almost the same)
\begin{align}
  \mathbb{E}_{t\sim [0,T]}\mathbb{E}_{\mathbf{x}(t)\sim p(\mathbf{x}(t))} 
  \left[ \frac{1}{2}\parallel \mathfrak{S}_{\theta}(\mathbf{x}(t),t,\mathbf{m}) -\nabla_{\mathbf{x}(t)}  \log  p(\mathbf{x}(t))\parallel^2\right].
  \label{eq:esm_loss}
\end{align}

Generally speaking, in the low-density data manifold area, the score estimation will be inaccurate, which will further lead to the low quality of the 
sampled data~\citep{song2019generative}. If the mel spectrogram or wave signal is contaminated with a very small scale noise, then the contaminated mel spectrogram or 
wave signal will spread to the 
entire space $\mathbb{R}^d$ instead of being limited to a small low-dimensional manifold. When using perturbed mel spectrogram or wave signal as input,
 the following denoising score matching (DSM) loss~\citep{vincent2011connection,song2019generative} 
\begin{align}
  \text{DSM loss}=\mathbb{E}_{t\sim [0,T]}\mathbb{E}_{\mathbf{x}(0)\sim p_{mel}(\mathbf{x}(0))} \mathbb{E}_{\mathbf{x}(t)\sim p(\mathbf{x}(t)|\mathbf{x}(0))} 
  \left[ \frac{1}{2}\parallel \mathfrak{S}_{\theta}(\mathbf{x}(t),t,\mathbf{m}) -\nabla_{\mathbf{x}(t)}  \log  p(\mathbf{x}(t)|\mathbf{x}(0))\parallel^2 \right].
  \label{eq:dsm_loss}
\end{align}
is equal to the loss~(\ref{eq:esm_loss}) of a non-parametric (e.g. Parzen windows densiy) estimator~\citep{vincent2011connection}.
This DSM loss is used in this paper to train the score prediction network.
If we can accurately estimate the score $\nabla_{\mathbf{x}(t)}  \log  p(\mathbf{x}(t))$ of the distribution, then we can generate 
mel spectrogram  or wave sample data from the original distribution.  

It should be \textbf{noted} that in the experiment we found that the choice of training loss is very critical. For It\^oTTS, 
the $\mathcal{L}1$ loss is more suitable, and for It\^oWave, the $\mathcal{L}2$ loss is more appropriate. 
The intuition is that the frequency mel-spectrogram data favors $\mathcal{L}1$-loss, while time-domain wave data favors $\mathcal{L}2$-loss.

Generally the transition densities $p(\mathbf{x}(t)|\mathbf{x}(0))$ and $\nabla_{\mathbf{x}(t)}  \log  p(\mathbf{x}(t)|\mathbf{x}(0))$ 
in the DSM loss are difficult to calculate, but for linear SDE, these values 
have close formulas~\citep{sarkka2019applied}.

\subsection{Linear SDE and transition densities}
\label{sec:linear_sde}

Linear SDE refers to such an equation
\begin{align}
  \left\{
  \begin{array}{lr}
 d\mathbf{X}=\left[\mathbf{C}(t)\mathbf{X}+\mathbf{d}(t)\right]dt+g(t)\mathbf{I}d\mathbf{W} &  \\
 \mathbf{X}(0)=\mathbf{x}(0). &
\end{array}
\right.
 \label{eq:lineare_ito_sde}
\end{align}
where $\mathbf{C}(t)\in\mathbb{R}^{d\times d}$ and $\mathbf{d}(t)\in\mathbb{R}^d$.

The transition densities $p(\mathbf{x}(t)|\mathbf{x}(0))$ of the solution process $\mathbf{X}(t)$ for the SDE~(\ref{eq:lineare_ito_sde}) 
is the solution to the Fokker-Planck-Kolmogorov (FPK)
equation~\citep{sarkka2019applied}
\begin{align}
 \frac{\partial p(\mathbf{x}(t),t)}{\partial t}  = -\sum_{i=1}^d\frac{\partial \left[ \mathbf{C}(t)\mathbf{x}+\mathbf{d}(t)\right]}{\partial x_i} 
 +\sum_{i=1}^d\sum_{j=1}^d \frac{\partial^2 }{\partial x_i \partial x_j } [g(t)^2p(\mathbf{x}(t),t)],
 \label{eq:fpk_pde}
\end{align}
which in this case can derive that $p(\mathbf{x}(t)|\mathbf{x}(0))$ is Gaussian with mean $\mathbf{m}(t)$ and variance
 $\mathbf{V}(t)$ satisfy the ordinary equations~\citep{sarkka2019applied}
\begin{align}
  \left\{
  \begin{array}{lr}
    \frac{d \mathbf{m}(t)}{dt} = \mathbf{C}(t)\mathbf{m}(t)+\mathbf{d}(t) &  \\
    \frac{d \mathbf{V}(t)}{dt} = \mathbf{C}(t)\mathbf{V}(t) + \mathbf{V}(t)\mathbf{C}(t)^T+g(t)^2\mathbf{I}.&
\end{array}
\right.
  \label{eq:mean_var_pde}
 \end{align}
Empirically it is found that different types of linear SDE for different audio generation tasks, e.g. the 
variance exploding (VE) SDE~\citep{song2020score} is much suitable for wave generation, while variance preserving (VP) SDE~\citep{song2020score} 
is more suitable mel-spectrogram generation.
VE SDE is of the following form
 \begin{align}
  \left\{
  \begin{array}{lr}
 d\mathbf{X}=\sigma_0(\frac{\sigma_1}{\sigma_0})^t\sqrt{2\log \frac{\sigma_1}{\sigma_0}}d\mathbf{W} &  \\
 \mathbf{X}(0)=\mathbf{x}(0)\sim \int p_{mel}(\mathbf{x})\mathcal{N}(\mathbf{x}(0);
 \mathbf{x},\sigma_0^2\mathbf{I})d\mathbf{x}, &
\end{array}
\right.
 \label{eq:ve_sde}
\end{align}
where $\sigma_0=0.01 < \sigma_1$.

Then the differential equation satisfied by the mean and variance of the transition densities $p(\mathbf{x}(t)|\mathbf{x}(0))$ is as follows
\begin{align}
  \left\{
  \begin{array}{lr}
    \frac{d \mathbf{m}(t)}{dt} = \mathbf{0} &  \\
    \frac{d \mathbf{V}(t)}{dt} = 2\sigma^2_0(\frac{\sigma_1}{\sigma_0})^{2t}\log \frac{\sigma_1}{\sigma_0}\mathbf{I}.&
\end{array}
\right.
  \label{eq:mean_var_pde_no1}
 \end{align}
 Solving the above equation, and choose $\sigma_1$ makes $2\log \frac{\sigma_1}{\sigma_0}=1$, we get
the transition density of this as
\begin{align}
p(\mathbf{x}(t)|\mathbf{x}(0))=\mathcal{N}\left(\mathbf{x}(t);
\mathbf{x}(0),\left[\sigma_0^2(\frac{\sigma_1}{\sigma_0})^{2t}-\sigma_0^2\right]\mathbf{I}\right).
 \label{eq:ve_sde_trans_density}
\end{align}

The score of the VE linear SDE is
\begin{align}
  &\nabla_{\mathbf{x}(t)}\log p(\mathbf{x}(t)|\mathbf{x}(0))=\nabla_{\mathbf{x}(t)} \log \mathcal{N}\left(\mathbf{x}(t); 
  \mathbf{x}(0),\left[\sigma_0^2(\frac{\sigma_1}{\sigma_0})^{2t}-\sigma_0^2\right]\mathbf{I}\right)  \nonumber \\&
  =\nabla_{\mathbf{x}(t)} \left[ -\frac{d}{2}\log \left[2\pi \left(\sigma_0^2(\frac{\sigma_1}{\sigma_0})^{2t}-\sigma_0^2\right)\right]- 
  \frac{\parallel\mathbf{x}(t) - \mathbf{x}(0)\parallel^2}{2 \left( \sigma_0^2(\frac{\sigma_1}{\sigma_0})^{2t}-\sigma_0^2  \right)}  \right]
  = -  \frac{\mathbf{x}(t) - \mathbf{x}(0)}{ \sigma_0^2(\frac{\sigma_1}{\sigma_0})^{2t}-\sigma_0^2}. 
   \label{eq:ve_sde_score}
  \end{align}

The prior distribution $p(\mathbf{x}(T))$ is a  Gaussian
\begin{align}
\mathcal{N}\left(\mathbf{x}(T);\mathbf{0},\sigma_1^2\mathbf{I}\right)=
\frac{\exp (-\frac{1}{2\sigma_1^2}\left\lVert \mathbf{x}(T)\right\rVert^2) }{\sigma_1^d\sqrt{(2\pi)^d}},
\label{eq:ve_sde_prior}
\end{align}
thus $\log p(\mathbf{x}(T)) = -\frac{d}{2}\log(2\pi\sigma_1^2)
  - \frac{1}{2\sigma_1^2}\left\lVert \mathbf{x}(T)\right\rVert^2 $.

  The VP linear SDE is
  \begin{align}
    \left\{
    \begin{array}{lr}
   d\mathbf{X}=-\frac{1}{2}(\beta_0+t(\beta_1-\beta_0))\mathbf{X}dt + \sqrt{\beta_0+t(\beta_1-\beta_0)}d\mathbf{W} &  \\
   \mathbf{X}(0)=\mathbf{x}(0)\sim p_{mel}(\mathbf{x}), &
  \end{array}
  \right.
   \label{eq:vp_sde}
  \end{align}
  where $\beta_1 > \beta_0$ are constant hyperparameters.
  Then the differential equation satisfied by the mean and variance of the transition densities $p(\mathbf{x}(t)|\mathbf{x}(0))$ is as follows
  \begin{align}
    \left\{
    \begin{array}{lr}
      \frac{d \mathbf{m}(t)}{dt} =  -\frac{1}{2}(\beta_0+t(\beta_1-\beta_0))\mathbf{m}(t)&  \\
      \frac{d \mathbf{V}(t)}{dt} = -(\beta_0+t(\beta_1-\beta_0))\mathbf{V}(t)  + (\beta_0+t(\beta_1-\beta_0))\mathbf{I}.&
  \end{array}
  \right.
    \label{eq:mean_var_pde_no2}
   \end{align}
  By solving the above linear ordinary differential equations with initial conditions that $\mathbf{m}(0)=\mathbf{x}(0)$ and $\mathbf{V}(0)=\mathbf{x}(0)$, we obtain
  \begin{align}
    \left\{
    \begin{array}{lr}
      \mathbf{m}(t) =   \mathbf{x}(0)\exp{\left[-\frac{1}{2}\beta_0t-\frac{1}{4}t^2(\beta_1-\beta_0)\right]}&  \\
      \mathbf{V}(t)= \mathbf{I}- \mathbf{I}\exp{\left[-\beta_0t-\frac{1}{2}t^2(\beta_1-\beta_0)\right]}.&
  \end{array}
  \right.
    \label{eq:mean_var_no2}
   \end{align}
  
   we get
  the transition density of this as
  \begin{align}
  p(\mathbf{x}(t)|\mathbf{x}(0))=\mathcal{N}\left(\mathbf{x}(t);
  \mathbf{x}(0)\exp{\left[-\frac{1}{2}\beta_0t-\frac{1}{4}t^2(\beta_1-\beta_0)\right]}, \mathbf{I}- \mathbf{I}\exp{\left[-\beta_0t-\frac{1}{2}t^2(\beta_1-\beta_0)\right]}\right).
   \label{eq:ve_sde_trans_density}
  \end{align}
  
  The score of the VP linear SDE is
  \begin{align}
    &\nabla_{\mathbf{x}(t)}\log p(\mathbf{x}(t)|\mathbf{x}(0)) \nonumber \\&
    =\nabla_{\mathbf{x}(t)} \left[ -\frac{d}{2}\log \left[2\pi \left(1-\exp{\left[-\beta_0t-\frac{1}{2}t^2(\beta_1-\beta_0)\right]}\right)\right]- 
    \frac{\parallel\mathbf{x}(t) - \mathbf{x}(0)\exp{\left[-\frac{1}{2}\beta_0t-\frac{1}{4}t^2(\beta_1-\beta_0)\right]}\parallel^2}{2 \left( 1-\exp{\left[-\beta_0t-\frac{1}{2}t^2(\beta_1-\beta_0)\right]}  \right)}  \right]
    \nonumber  \\& = -  \frac{\mathbf{x}(t) - \mathbf{x}(0)\exp{\left[-\frac{1}{2}\beta_0t-\frac{1}{4}t^2(\beta_1-\beta_0)\right]}}{1-\exp{\left[-\beta_0t-\frac{1}{2}t^2(\beta_1-\beta_0)\right]} }. 
     \label{eq:ve_sde_score}
    \end{align}
  
    The prior distribution $p(\mathbf{x}(T))$ is a  Gaussian
    \begin{align}
    \mathcal{N}\left(\mathbf{x}(T);\mathbf{0},\mathbf{I}\right)=
    \frac{\exp (-\frac{1}{2}\left\lVert \mathbf{x}(T)\right\rVert^2) }{\sqrt{(2\pi)^d}},
    \label{eq:vp_sde_prior}
    \end{align}
    thus $\log p(\mathbf{x}(T)) = -\frac{d}{2}\log(2\pi)
      - \frac{1}{2}\left\lVert \mathbf{x}(T)\right\rVert^2 $.

      In this paper, we use the VE linear SDE for wave generation and the VP linear SDE for mel-spectrogram generation.
  In future work, we will develop other types of linear SDEs that are suitable for audio generation, and discover the principles that can determine which type of SDE is 
  more appropriate for audio generation.

\subsection{Training algorithms}
\label{sec:training_algorithm}

Based on subsections~\ref{sec:dist_transormation_sde} and~\ref{sec:score_estimation}, we can get the training algorithm of the score networks based 
on general SDEs. Here we have 
taken the training algorithm of It\^oTTS as an example, as shown below
\begin{algorithm}[H]
  \caption{Training of the score network in general SDE-based mel spectrogram generation model}
  \label{alg:mel_sde_score_training}
  
  \textbf{Input and initialization}: The mel spectrogram $\mathbf{x}$  and the corresponding text condition $\mathbf{p}$, the diffusion time $T$.
  
  1: \textbf{for} $k=0, 1, \cdots$
  
  2:  \quad  Uniformly sample $t$ from $[0, T]$.

 3:  \quad \quad Randomly sample batch of $\mathbf{x}$ and  $\mathbf{p}$, let $\mathbf{x}(0)=\mathbf{x}$, compute to sample $\mathbf{x}(t)$ and 
 $\nabla_{\mathbf{x}(t)}  \log  p(\mathbf{x}(t)|\mathbf{x}(0))$, then average the following
 \begin{align}
  \text{DSM } loss=
\parallel \mathfrak{S}_{\theta_k}(\mathbf{x}(t),t,\mathbf{p}) -\nabla_{\mathbf{x}(t)}  \log  p(\mathbf{x}(t)|\mathbf{x}(0))\parallel_1. \nonumber
\end{align}
  
4:  \quad \quad Do the back-propagation and the parameter updating of $\mathfrak{S}_{\theta}$.

5: \quad $k \leftarrow k+1$.

 6: \textbf{Until} stopping conditions are satisfied and $\mathfrak{S}_{\theta_k}$ converges, e.g. to $\mathfrak{S}_{\theta_*}$.
  
  \textbf{Output}: $\mathfrak{S}_{\theta_*}$.
  \end{algorithm}

In the above subsection~\ref{sec:linear_sde}, for VE linear SDE, we have obtained the closed-form of the score, that is, we have obtained the training signal 
of the neural networks
$\mathfrak{S}_{\theta}(\mathbf{x}(t),t,\mathbf{p})$ and $\mathfrak{S}_{\theta}(\mathbf{x}(t),t,\mathbf{m})$. 
At this point, we can train these score estimation networks using Algorithm~\ref{alg:mel_sde_score_training} with step 3 modified as  

      \fbox{%
      \parbox{\textwidth}{%
          3:  \quad \quad Randomly sample batch of $\mathbf{x}$ and  $\mathbf{p}$, let $\mathbf{x}(0)=\mathbf{x}$. Sample $\mathbf{x}(t)$ from the 
          distribution $\mathcal{N}\left(\mathbf{x}(t);
          \mathbf{x}(0),\left[\sigma_0^2(\frac{\sigma_1}{\sigma_0})^{2t}-\sigma_0^2\right]\mathbf{I}\right)$, compute the target score 
          as $-  \frac{\mathbf{x}(t) - \mathbf{x}(0)}{ \sigma_0^2(\frac{\sigma_1}{\sigma_0})^{2t}-\sigma_0^2 } $. Average the following
          \begin{align}
           \text{DSM } loss=
          \parallel \mathfrak{S}_{\theta_k}(\mathbf{x}(t),t,\mathbf{p}) +  \frac{\mathbf{x}(t) - \mathbf{x}(0)}{ \sigma_0^2(\frac{\sigma_1}{\sigma_0})^{2t}-\sigma_0^2 }\parallel_1 . \nonumber
         \end{align} 
      }%
    }

  For VP linear SDE, we have obtained the closed-form of the score, that is, we have obtained the training signal 
  of the neural networks
  $\mathfrak{S}_{\theta}(\mathbf{x}(t),t,\mathbf{p})$ and $\mathfrak{S}_{\theta}(\mathbf{x}(t),t,\mathbf{m})$. 
  At this point, we can train these score estimation networks using Algorithm~\ref{alg:mel_sde_score_training} with step 3 modified as  
  
        \fbox{%
        \parbox{\textwidth}{%
            3:  \quad \quad Randomly sample batch of $\mathbf{x}$ and  $\mathbf{p}$, let $\mathbf{x}(0)=\mathbf{x}$. Sample $\mathbf{x}(t)$ from the 
            distribution 
            $\mathcal{N}\left(\mathbf{x}(t);
            \mathbf{x}(0)\exp{\left[-\frac{1}{2}\beta_0t-\frac{1}{4}t^2(\beta_1-\beta_0)\right]}, 
            \mathbf{I}- \mathbf{I}\exp{\left[-\beta_0t-\frac{1}{2}t^2(\beta_1-\beta_0)\right]}\right)$, compute the target score 
            as $-  \frac{\mathbf{x}(t) - \mathbf{x}(0)\exp{\left[-\frac{1}{2}\beta_0t-\frac{1}{4}t^2(\beta_1-\beta_0)\right]}}{1-\exp{\left[-\beta_0t-\frac{1}{2}t^2(\beta_1-\beta_0)\right]} }$. 
            Average the following
            \begin{align}
             \text{DSM } loss=
            \parallel \mathfrak{S}_{\theta_k}(\mathbf{x}(t),t,\mathbf{p}) +  \frac{\mathbf{x}(t) - \mathbf{x}(0)\exp{\left[-\frac{1}{2}\beta_0t-\frac{1}{4}t^2(\beta_1-\beta_0)\right]}}{1-\exp{\left[-\beta_0t-\frac{1}{2}t^2(\beta_1-\beta_0)\right]} }\parallel_1 . \nonumber
           \end{align} 
        }%
      }

\subsection{Mel spectrogram sampling and wave sampling}
\label{sec:sampling}

After we get the optimal score network  $\mathfrak{S}_{\theta_*}$ through loss minimization, thus we can get the gradient of log value of the distribution 
probability density of the mel spectrogram or wave with $\mathfrak{S}_{\theta_*}(\mathbf{x}(t),t,\mathbf{p})$  
or $\mathfrak{S}_{\theta_*}(\mathbf{x}(t),t,\mathbf{m})$.
Then we can use Langevin dynamics or the reverse-time It\^o SDE~(\ref{eq:reverse_ito_sde})
to generate the mel spectrogram corresponding to the specific text $\mathbf{p}$ or the wave corresponding to the specific mel spectrogram $\mathbf{m}$.
The reverse-time SDE~(\ref{eq:reverse_ito_sde}) 
can be solved and used to generate target audio data.
Assuming that the time schedule is fixed, the discretization of the diffusion process~(\ref{eq:ito_sde}) is as follows
\begin{align}
  \left\{
  \begin{array}{lr}
 \mathbf{X}(i\Delta t+\Delta t) - \mathbf{X}(i\Delta t) =\mathbf{f}(\mathbf{X}(i\Delta t),i\Delta t)\Delta t+g(i\Delta t)\mathbf{\xi}(i\Delta t) \quad (i=0,1,\cdots, N-1)&  \\
 \mathbf{X}(0)=\mathbf{x}(0) &
\end{array}
\right.
 \label{eq:discret_ito_sde}
\end{align}
since $d\mathbf{W}$ is a wide sense stationary white noise process~\citep{evans2012introduction}, which is denoted as $\mathbf{\xi}(\cdot)\sim 
\mathcal{N}\left(\mathbf{0},\mathbf{I}\right)$ in this paper.

The corresponding discretization of the reverse-time diffusion process~(\ref{eq:reverse_ito_sde}) is 
\begin{align}
  \left\{
  \begin{array}{lr}
    \mathbf{X}(i\Delta t) - \mathbf{X}(i\Delta t+\Delta t) =\mathbf{f}(\mathbf{X}(i\Delta t+\Delta t),i\Delta t+\Delta t)  (-\Delta t) &  \\
    -g(i\Delta t+\Delta t)^2 \mathfrak{S}_{\theta_*}(\mathbf{X}(i\Delta t+\Delta t),i\Delta t+\Delta t, \textbf{m}) (-\Delta t) + g(i\Delta t+\Delta t)\mathbf{\xi}(i\Delta t) &  \\
 \mathbf{X}(T)=\mathbf{x}(T)& \\
 T = N\Delta t, \quad i=0,1,\cdots, N-1 &
\end{array}
\right.
 \label{eq:discret_reverse_ito_sde}
\end{align}

In this paper, we use the strategy of~\citet{song2020score}, which means that at each time step,  Langevin dynamics is used to predict first, and then reverse-time 
It\^o SDE~(\ref{eq:discret_reverse_ito_sde}) is used to revises the first predicted result.

The generation algorithm of mel spectrogram based on general linear SDE is as follows
\begin{algorithm}[H]
  \caption{General SDE-based It\^oTTS mel spectrogram generation algorithm}
  \label{alg:general_sde_sampling}
  
  \textbf{Input and initialization}: the score network $\mathfrak{S}_{\theta_*}$, input text $\mathbf{p}$, and $\mathbf{x}( N\Delta t)\sim 
  \mathcal{N}\left(\mathbf{0},s\mathbf{I}\right)$.
  
  1: \textbf{for} $k=N-1, \cdots, 0$
  
  2:  \quad   $\mathbf{x}(k\Delta t) = \mathbf{x}(k\Delta t+\Delta t) - \mathbf{f}(\mathbf{x}(k\Delta t+\Delta t),k\Delta t+\Delta t)  \Delta t
  +g(k\Delta t+\Delta t)^2 \mathfrak{S}_{\theta_*}(\mathbf{x}(k\Delta t+\Delta t),k\Delta t+\Delta t, \mathbf{p}) \Delta t + g(k\Delta t+\Delta t)\mathbf{\xi}(k\Delta t) $

 3: \quad  $\mathbf{x}(k\Delta t) \leftarrow \mathbf{x}(k\Delta t) + \epsilon_k \mathfrak{S}_{\theta_*}
 (\mathbf{x}(k\Delta t),k\Delta t, \mathbf{p})  +\sqrt{2\epsilon_k}\mathbf{\xi}(k\Delta t)$
  
  4: \quad $k \leftarrow k-1$.
  
  \textbf{Output}: The generated mel spectrogram $\mathbf{x}(0)$.
  \end{algorithm}

  The generation algorithm of mel spectrogram based on VP linear SDE is same as Algorithm~\ref{alg:general_sde_sampling} with step 2 modified as follows
  \fbox{%
  \parbox{\textwidth}{%
    2:  \quad   $\mathbf{x}(k\Delta t) = \mathbf{x}(k\Delta t+\Delta t) 
    + 2\sigma_0^2(\frac{\sigma_1}{\sigma_0})^{2k\Delta t+2\Delta t}\log \frac{\sigma_1}{\sigma_0}
    \mathfrak{S}_{\theta_*}(\mathbf{x}(k\Delta t+\Delta t),k\Delta t+\Delta t, \mathbf{p}) \Delta t + 
    \sigma_0(\frac{\sigma_1}{\sigma_0})^{k\Delta t+\Delta t}\sqrt{2\log \frac{\sigma_1}{\sigma_0}} \mathbf{\xi}(k\Delta t) $
   }%
   }

The wave sampling algorithm in It\^oWave is similar, please refer to the appendix.

\subsection{Architectures of  $\mathfrak{S}_{\theta}(\mathbf{x}(t),t,\mathbf{p})$  
and $\mathfrak{S}_{\theta}(\mathbf{x}(t),t,\mathbf{m})$.}
\label{sec:architecture}

It was found that although the score network model does not have as strict restrictions 
on the network structure as the flow model~\citep{miao2020flow,prenger2019waveglow,valle2020flowtron}, not all network structures are suitable for score prediction.

The structure of \textbf{It\^oTTS}'s score prediction network  $\mathfrak{S}_{\theta}(\mathbf{x}(t),t,\mathbf{p})$  is using an encoder-decoder framework, as shown 
in Figure~\ref{itowave_arch}. Among them, the encoder is borrowed from Fastspeech2~\citep{ren2020fastspeech}. Its function is to encode text (text will first become 
the sequence 
of phonemes by using the text cleaner from https://github.com/keithito/tacotron), and then expand to the actual length of mel spectrogram. The encoder first encodes 
the text that contains the 
sinusoid position encoding information, and then sends the encoding to the $N$-layer feed-forward Transformer (FFT) block to obtain the feature map of the encoded 
phonemes, which will be used to predict the  duration, pitch, and energy of the phonemes. Among them, the pitch and energy information will be added to the feature map of the 
phonemes after embedding. The duration is used to extend the feature map to the actual length of the mel spectrogram. The feature map will be sent to the decoder as 
the conditional input of text. 

The \textbf{decoder} of It\^oTTS has two other inputs, one is the conditional step time information, the other is the mel spectrogram.
The mel spectrogram will go through a convolution level and several 
linear layers and Sigmoid linear unit (SILU)~\citep{elfwing2018sigmoid} layers for encoding; the step time information will go through a 
Gaussian Fourier projection (GFP)~\citep{ren2020fastspeech} module, and the output encoding will 
be added to the encoded feature map of mel spectrogram. The sum will be sent to the key module of the decoder, 
which consists of several dilated residual blocks. 
Each dilated residual block has two convolution layers and two CHUNK layers, whose function is to evenly divide the feature map into two parts.
That means each dilated residual block  has two outputs, one is the status information, which is sent to the next residual block, and the other is part 
of the score information as output. The output of the encoder, which is the text information, will be sent to each residual block, and be the main conditional 
variable, which controls 
the output of the decoder. Finally, the output of all residual blocks is averaged, and then after two convolution layers, the final score is obtained.

In the experiments, we found that if only one decoder is used to predict the scores of all channels of the mel spectrogram at once, the final synthesized mel spectrogram
always has some channels, especially the low-frequency part that cannot be generated.  
Moreover, it is found that the most accurate scores are concentrated in high-frequency channels that do not have many harmonic textures, which seems to be easy to learn.
On the contrary, if multiple decoders are used, and each decoder is only responsible for few channels of the mel coefficients, then perfect results can be obtained.
For details of this observation, please refer to the appendix.

\begin{figure}[th]
  \centering
  \hspace{-5mm}
  \includegraphics[width=1.0\linewidth]{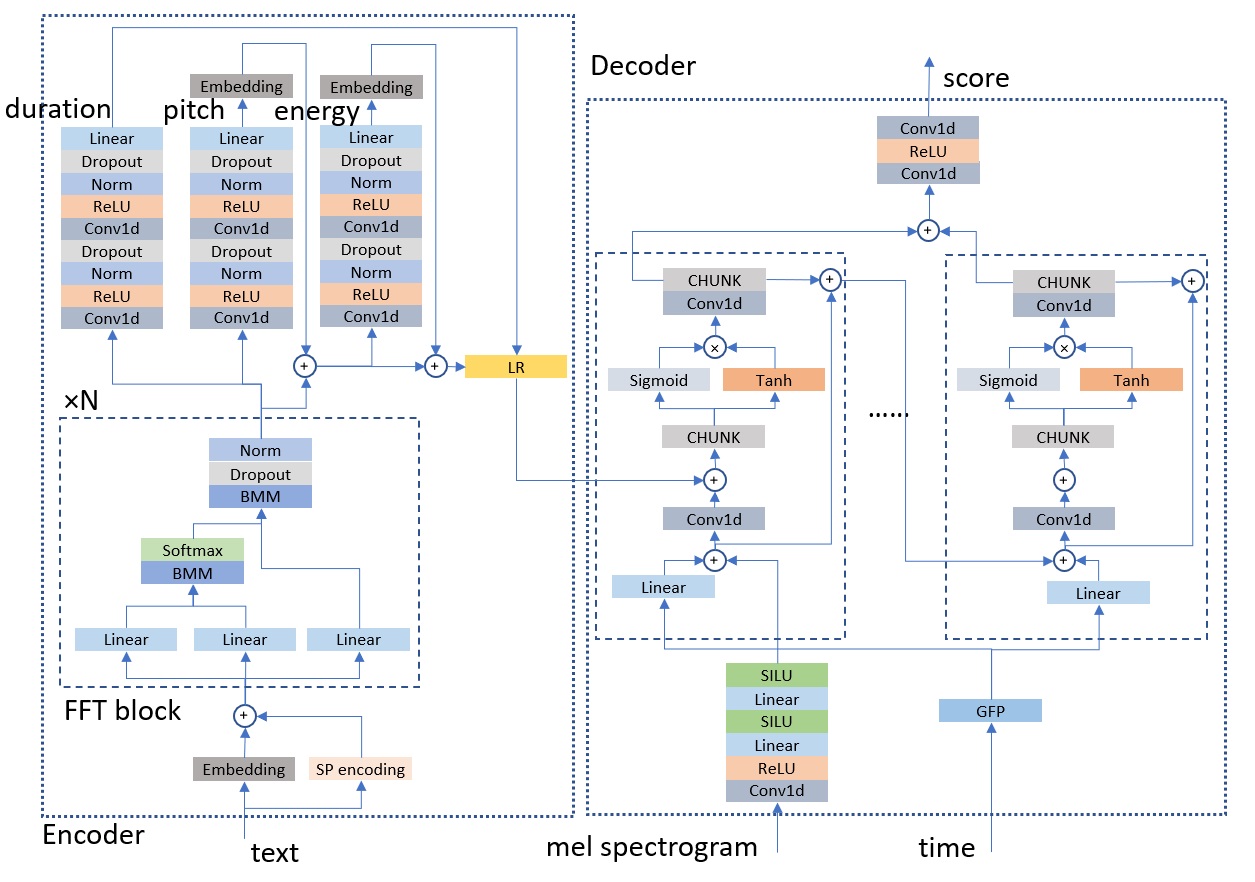}
  \hspace{-5mm}
\caption{
The architecture of It\^oTTS.
    }
  \label{itowave_arch}
  \end{figure}

  \textbf{It\^oWave}'s score prediction network structure $\mathfrak{S}_{\theta}(\mathbf{x}(t),t,\mathbf{m})$,  as shown in Figure~\ref{itowave_arch}, is slightly simpler 
  than $\mathfrak{S}_{\theta}(\mathbf{x}(t),t,\mathbf{p})$. The input is the wave to be generated, and the conditional input has mel spectrogram and time $t$. 
  The output is the score at time $t$. All three types of input require preprocessing processes. The preprocessing of the wave is through a 
  convolution layer; the preprocessing of mel spectrogram is based on the upsampling by two  transposed convolution layers; and the preprocessing of step time $t$ 
  is the same as in \textbf{It\^oTTS}. After all inputs are preprocessed, they will be sent to the 
  most critical module of $\mathfrak{S}_{\theta}(\mathbf{x}(t),t,\mathbf{m})$, which is very similar to the decoder 
  of $\mathfrak{S}_{\theta}(\mathbf{x}(t),t,\mathbf{p})$ in It\^oTTS, that is several serially connected dilated residual blocks. The main input 
  of the dilated residual block is the wave, and the step time condition and mel spectrogram condition will be input into these dilated residual blocks 
  one after another, and added to the feature map after the transformation of the wave signal. Similarly, there are two outputs of each dilated 
  residual block, one is the state, which is used for input to the next residual block, and the other is the final output. The 
  advantage of this is the ability to synthesize information of different granularities. Finally, the outputs of all residual 
  blocks are summed and then pass through two convolution layers as the final output score.

  \begin{figure}[th]
    \centering
    \hspace{-5mm}
    \includegraphics[width=1.0\linewidth]{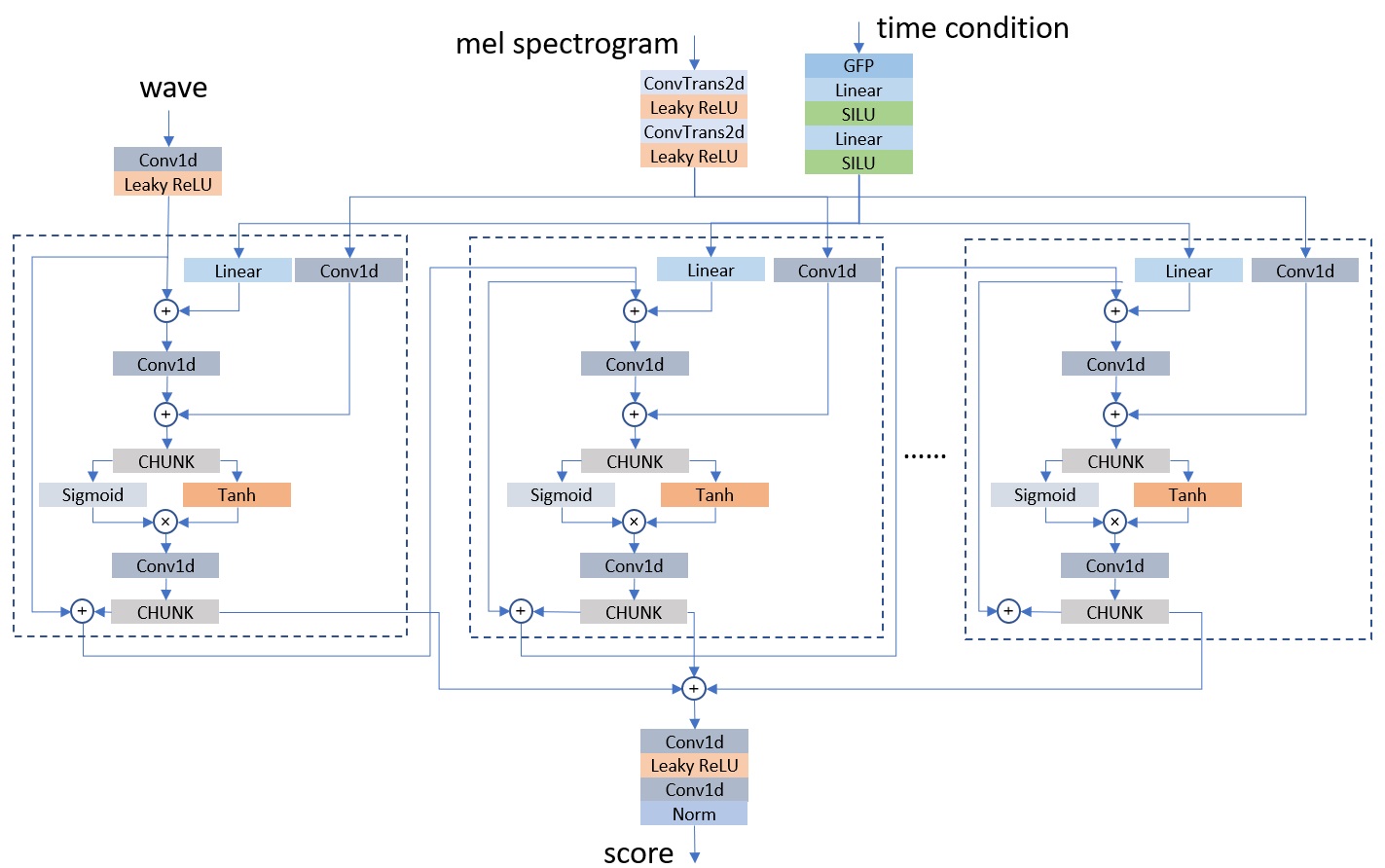}
    \hspace{-5mm}
  \caption{
  The architecture of It\^oWave.
      }
    \label{itowave_arch}
    \end{figure}

\section{Experiments}
\label{sec:experiments}


\subsection{Dataset and setup}

The data set we use is LJSpeech~\citep{ito2017lj}, a single female speech database, with a total of 24 hours, 13100 sentences, randomly divided into 13000/50/50 for 
training/verification/testing. The sampling rate is 22050.  In the experiment, all the places that involve mel spectrogram, whether it is in It\^oTTS or 
in It\^oWave, window length is 1024, hop length is 256, the number of mel channels is 80. 
We use the same Adam~\citep{kingma2014adam} training algorithm for It\^oTTS and It\^oWave. We have done 
quantitative evaluations based on mean opinion score (MOS) 
with other state-of-the-art methods on It\^oTTS and It\^oWave respectively. For It\^oTTS, 
we compared with Tacotron 2~\citep{shen2018natural} and Fastspeech 2~\citep{ren2020fastspeech}, and for It\^oWave, we 
compared with WaveNet~\citep{oord2016wavenet}, WaveGlow~\citep{prenger2019waveglow}, Diffwave~\citep{kong2020diffwave}, and WaveGrad~\citep{chen2020wavegrad}.
All experiments were performed on a GeForce RTX 3090 GPU with 24G memory.

\subsection{Results and discussion}

In order to verify the naturalness and fidelity of the synthesized voice, we randomly select 40  from 50 test data for each subject, 
and then let the subject give the synthesized sound a MOS score of 0-5. 

In order to compare the synthesis quality of It\^oTTS with Tacotron 2~\citep{shen2018natural} and Fastspeech 2~\citep{ren2020fastspeech}, 
a pre-trained HiFi-GAN model~\citep{su2020hifi} as a vocoder to transform 
the mel spectrogram into a wave. In the experiment, It\^oTTS uses 4 FFT layers in the encoder, three variance adaptors are used to predict the duration, 
pitch and energy of the phoneme respectively. 8 decoders, each of which is responsible for 10 channels of the mel spectrogram. 
Each decoder has 30 residual layers. The parameters of VE linear 
SDE in the experiment are $\sigma_0=0.01$, $\sigma_1=50$, and the number of time steps $N=1000$. The parameters of SDE used in It\^oWave are the same.

In the structure of Tacotron 2~\citep{shen2018natural}, there are 3 convolution layers in the encoder, the dimensions of recurrent neural network (RNN) and 
attention in the decoder are both 1024, and the 
postnet has 5 convolution layers. The batch size is 32.
In the structure of Fastspeech 2~\citep{ren2020fastspeech}, the FFT block in the encoder has 4 layers, the FFT block in the decoder has 6 layers, and the dropout in both the encoder 
and the decoder is 0.2.

The results are shown in Table~\ref{tab:tts}. MOS with 95\% confidence is used in a comparative study of different state-of-the-art TTS systems on the 
test set of LJSpeech dataset. It can be seen that the MOS of It\^oTTS is slightly better than the previous state-of-the-art method.

\begin{table}[th]
  \caption[mos_tts]{MOS with 95\% confidence  in a comparative study of different state-of-the-art TTS methods on the 
  test set of LJspeech dataset. All methods use a pre-trained HiFi-GAN model as the vocoder.}\label{tab:tts}
  \centering
  \begin{tabular}{c|c}
  \hline
  \hline
  Methods & MOS\\
  \hline
  \hline
  Ground truth &4.45$\pm$ 0.07\\
  Tacotron 2 &3.775$\pm$  0.161 \\
  Fastspeech 2 &3.9$\pm$  0.159\\
  \hline
  \hline
  It\^oTTS   & 3.925$\pm$ 0.160 \\
  \hline
  \end{tabular}
  \end{table}


For the vocoder It\^oWave,  the original mel spectrogram of the test set was used as the condition input to the score estimation network.
It\^oWave uses 30 residual layers.
For the comparison methods, 
WaveNet~\citep{oord2016wavenet} uses 4 stacks with each stack consists of 24 dilated convolution layers.
WaveGlow~\citep{prenger2019waveglow} uses 12 flows, and each flow uses an 8-layer wavenet to similarly make a reversible transform.
Diffwave~\citep{kong2020diffwave} uses a 30-layer dilated residual block, the number of reesidual channels is 64.
WaveGrad~\citep{chen2020wavegrad} uses 5 layers of upsampling and 4 layers of downsampling.

The results are shown in Table~\ref{tab:vocoder}, and you can see that It\^oWave scores the best. 
It has approached the true value of ground truth. 

\begin{table}[th]
  \caption[mos_vocoder]{MOS with 95\% confience  in a comparative study of different state-of-the-art vocoders on the 
  test set of LJspeech dataset.}\label{tab:vocoder}
  \centering
  \begin{tabular}{c|c}
  \hline
  \hline
  Methods & MOS\\
  \hline
  \hline
  Ground truth &4.45$\pm$ 0.07\\
  WaveNet &4.3$\pm$  0.130 \\
 WaveGlow &3.95$\pm$  0.161\\
 DiffWave &4.325$\pm$  0.123\\
 WaveGrad &4.1$\pm$  0.158\\
  \hline
  \hline
  It\^oWave  & 4.35$\pm$ 0.115 \\
  \hline
  \end{tabular}
  \end{table}

\section{Conclusion}
\label{sec:conclusion}

This paper proposes   It\^oTTS and   It\^oWave, general methods based on linear SDE that can accomplish TTS and vocoder tasks at the same time. 
Under conditional input, It\^oTTS and   It\^oWave can continuously transform simple distributions into corresponding mel spectrogram data or wave 
data through reverse-time linear SDE and Langevin dynamic. It\^oTTS and   It\^oWave 
use neural networks to predict the required score for reverse-time 
linear SDE and Langevin dynamic sampling, which is the gradient of the log probability density at a specific time. For It\^oTTS and   It\^oWave, 
we designed the corresponding effective score prediction networks. Experiments show that the MOS of It\^oTTS and   It\^oWave 
can achieved the state-of-the-art respectively.

For future work, we believe that there are three important research directions. The first is to study different linear SDEs, that is, how different drift and 
diffusion coefficients affect the generation effect, and further, how to choose drift and diffusion coefficients to get the best Generation effect; 
in addition, study how to sample faster under the already trained score network, which is to speed up the generation speed; finally, extend this method 
 to the generation of discrete data, such as the generation of midi music.

\bibliographystyle{named}
\bibliography{ito_audio}



\appendix

\section{Appendix}
\label{sec:appendix}

This appendix contains a lot of details that are not expanded in the text, including  training algorithm of It\^oWave's score prediction 
network, It\^oWave's waveform generation algorithm, the influence of different numbers of  decoders in It\^oTTS, and examples 
of the generation process of mel spectrogram and wave in It\^oTTS and It\^oWave.




\subsection{Algorithms}
\label{sec:app_algorithms}

The training algorithms of the score network $\mathfrak{S}_{\theta}$ for It\^oWave  are  shown in Algorithm~\ref{alg:sde_wave_score_training} 
and Algorithm~\ref{alg:type1_sde_wave_score_training}.
\begin{algorithm}[H]
  \caption{Training of the score network in general SDE-based wave generation model}
  \label{alg:sde_wave_score_training}
  
  \textbf{Input and initialization}: The audio wave $\mathbf{x}$ and the corresponding mel spectrograms $\mathbf{m}$, the diffusion time $T$.
  
  1: \textbf{for} $k=0, 1, \cdots$
  
  2:  \quad  Uniformly sample $t$ from $[0, T]$.

 3:  \quad \quad Randomly sample batch of $\mathbf{x}$ and  $\mathbf{m}$, let $\mathbf{x}(0)=\mathbf{x}$, compute $\mathbf{x}(t)$ and 
 $\nabla_{\mathbf{x}(t)}  \log  p(\mathbf{x}(t)|\mathbf{x}(0))$, then average the following
 \begin{align}
  loss=
  \frac{1}{2}\parallel \mathfrak{S}_{\theta_k}(\mathbf{x}(t),t,\mathbf{m}) -\nabla_{\mathbf{x}(t)}  \log  p(\mathbf{x}(t)|\mathbf{x}(0))\parallel^2 . \nonumber
\end{align}
  
4:  \quad \quad Do the back-propagation and the parameter updating of $\mathfrak{S}_{\theta}$.

5: \quad $k \leftarrow k+1$.

 6: \textbf{Until} stopping conditions are satisfied and $\mathfrak{S}_{\theta_k}$ converges, e.g. to $\mathfrak{S}_{\theta_*}$.
  
  \textbf{Output}: $\mathfrak{S}_{\theta_*}$.
  \end{algorithm}

\begin{algorithm}[H]
  \caption{Training of the score network in VE SDE-based wave generation model}
  \label{alg:type1_sde_wave_score_training}
  
  \textbf{Input and initialization}: The audio wave $\mathbf{x}$ and the corresponding mel spectrograms $\mathbf{m}$, the diffusion time $T$.
  
  1: \textbf{for} $k=0, 1, \cdots$
  
  2:  \quad  Uniformly sample $t$ from $[0, T]$.

  3:  \quad \quad Randomly sample batch of $\mathbf{x}$ and  $\mathbf{m}$, let $\mathbf{x}(0)=\mathbf{x}$. Sample $\mathbf{x}(t)$ from the 
  distribution $\mathcal{N}\left(\mathbf{x}(t);
  \mathbf{x}(0),\left[\sigma_0^2(\frac{\sigma_1}{\sigma_0})^{2t}-\sigma_0^2\right]\mathbf{I}\right)$, compute the target score 
  as $-  \frac{\mathbf{x}(t) - \mathbf{x}(0)}{ \sigma_0^2(\frac{\sigma_1}{\sigma_0})^{2t}-\sigma_0^2 } $. Average the following
  \begin{align}
   \text{DSM } loss=
  \parallel \mathfrak{S}_{\theta_k}(\mathbf{x}(t),t,\mathbf{m}) +  \frac{\mathbf{x}(t) - 
  \mathbf{x}(0)}{ \sigma_0^2(\frac{\sigma_1}{\sigma_0})^{2t}-\sigma_0^2 }\parallel_1 . \nonumber
 \end{align} 
  
4:  \quad \quad Do the back-propagation and the parameter updating of $\mathfrak{S}_{\theta}$.

5: \quad $k \leftarrow k+1$.

 6: \textbf{Until} stopping conditions are satisfied and $\mathfrak{S}_{\theta_k}$ converges, e.g. to $\mathfrak{S}_{\theta_*}$.
  
  \textbf{Output}: $\mathfrak{S}_{\theta_*}$.
  \end{algorithm}

The wave generation algorithms of  It\^oWave  are  shown in Algorithm~\ref{alg:general_sde_wave_sampling} and Algorithm~\ref{alg:type1_sde_wave_sampling}.

  \begin{algorithm}[H]
    \caption{General SDE-based It\^oWave wave generation algorithm}
    \label{alg:general_sde_wave_sampling}
    
    \textbf{Input and initialization}: the score network $\mathfrak{S}_{\theta_*}$, input mel spectrogram $\mathbf{m}$, and $\mathbf{x}( N\Delta t)\sim 
    \mathcal{N}\left(\mathbf{0},s\mathbf{I}\right)$.
    
    1: \textbf{for} $k=N-1, \cdots, 0$
    
    2:  \quad   $\mathbf{x}(k\Delta t) = \mathbf{x}(k\Delta t+\Delta t) - \mathbf{f}(\mathbf{x}(k\Delta t+\Delta t),k\Delta t+\Delta t)  \Delta t
    +g(k\Delta t+\Delta t)^2 \mathfrak{S}_{\theta_*}(\mathbf{x}(k\Delta t+\Delta t),k\Delta t+\Delta t, \mathbf{m}) \Delta t + g(k\Delta t+\Delta t)\mathbf{\xi}(k\Delta t) $
  
   3: \quad  $\mathbf{x}(k\Delta t) \leftarrow \mathbf{x}(k\Delta t) + \epsilon_k \mathfrak{S}_{\theta_*}
   (\mathbf{x}(k\Delta t),k\Delta t, \mathbf{m})  +\sqrt{2\epsilon_k}\mathbf{\xi}(k\Delta t)$
    
    4: \quad $k \leftarrow k-1$.
    
    \textbf{Output}: The generated wave $\mathbf{x}(0)$.
    \end{algorithm}

  \begin{algorithm}[H]
    \caption{VE SDE-based It\^oWave wave generation algorithm}
    \label{alg:type1_sde_wave_sampling}
    
    \textbf{Input and initialization}: the score network $\mathfrak{S}_{\theta_*}$, input mel spectrogram $\mathbf{m}$, and $\mathbf{x}( N\Delta t)\sim 
    \mathcal{N}\left(\mathbf{0},s\mathbf{I}\right)$.
    
    1: \textbf{for} $k=N-1, \cdots, 0$
    
    2:  \quad   $\mathbf{x}(k\Delta t) = \mathbf{x}(k\Delta t+\Delta t) 
    + 2\sigma_0^2(\frac{\sigma_1}{\sigma_0})^{2k\Delta t+2\Delta t}\log \frac{\sigma_1}{\sigma_0}
    \mathfrak{S}_{\theta_*}(\mathbf{x}(k\Delta t+\Delta t),k\Delta t+\Delta t, \mathbf{m}) \Delta t + 
    \sigma_0(\frac{\sigma_1}{\sigma_0})^{k\Delta t+\Delta t}\sqrt{2\log \frac{\sigma_1}{\sigma_0}} \mathbf{\xi}(k\Delta t) $

   3: \quad  $\mathbf{x}(k\Delta t) \leftarrow \mathbf{x}(k\Delta t) + \epsilon_k \mathfrak{S}_{\theta_*}
   (\mathbf{x}(k\Delta t),k\Delta t, \mathbf{m})  +\sqrt{2\epsilon_k}\mathbf{\xi}(k\Delta t)$
    
    4: \quad $k \leftarrow k-1$.
    
    \textbf{Output}: The generated wave $\mathbf{x}(0)$.
    \end{algorithm}

\subsection{Decoding multiple subbands of mel spectrogram separately in It\^oTTS}

\begin{figure}[H]
  \centering
  \subfigure[Use the checkpoint obtained in the tenth epoch.]{
  \begin{minipage}[t]{0.33\linewidth}
  \centering
  \includegraphics[width=2.0in]{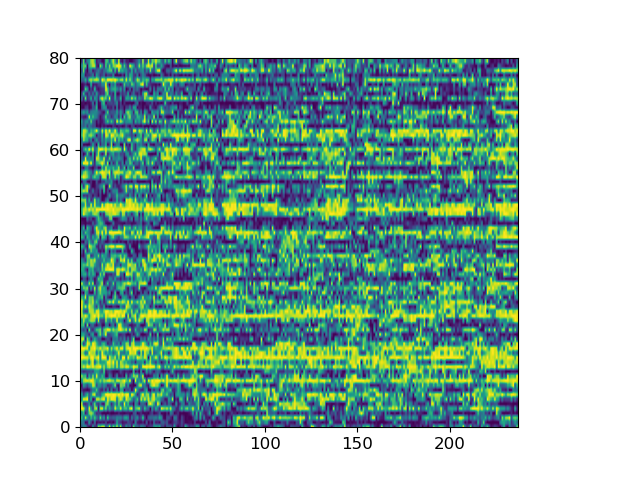}
  \label{tobe_single_band_tobe1000}
  \end{minipage}%
  }%
  \subfigure[In the 20th epoch.]{
  \begin{minipage}[t]{0.33\linewidth}
  \centering
  \includegraphics[width=2.0in]{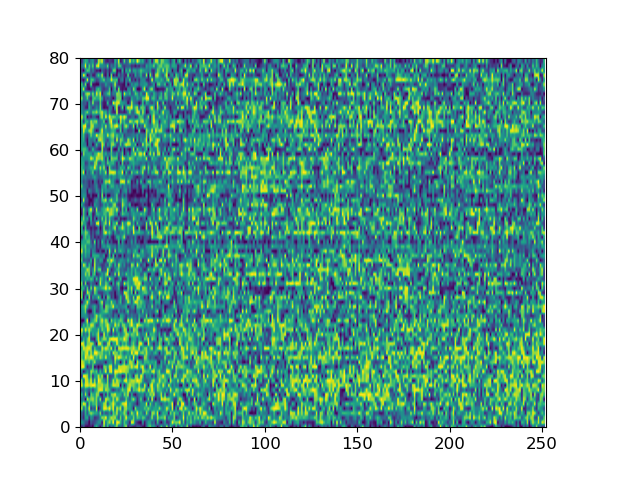}
  \label{tobe_single_band_tobe2000}
  \end{minipage}%
  }%
  \subfigure[In the 30th epoch.]{
    \begin{minipage}[t]{0.33\linewidth}
    \centering
    \includegraphics[width=2.0in]{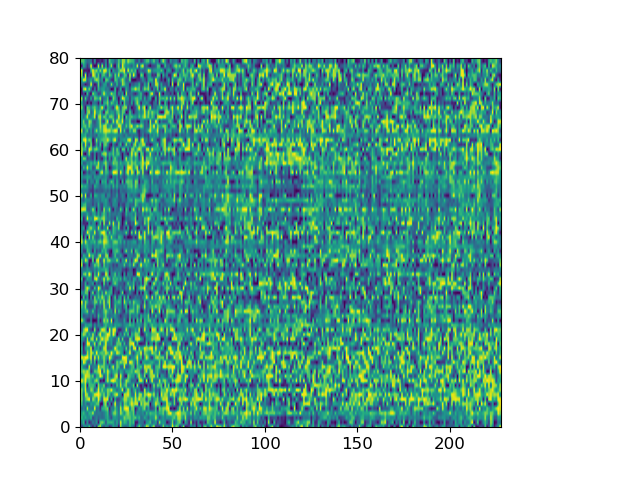}
    \label{tobe_single_band_tobe3000}
    \end{minipage}%
    }%

    \subfigure[In the 50th epoch.]{
      \begin{minipage}[t]{0.33\linewidth}
      \centering
      \includegraphics[width=2.0in]{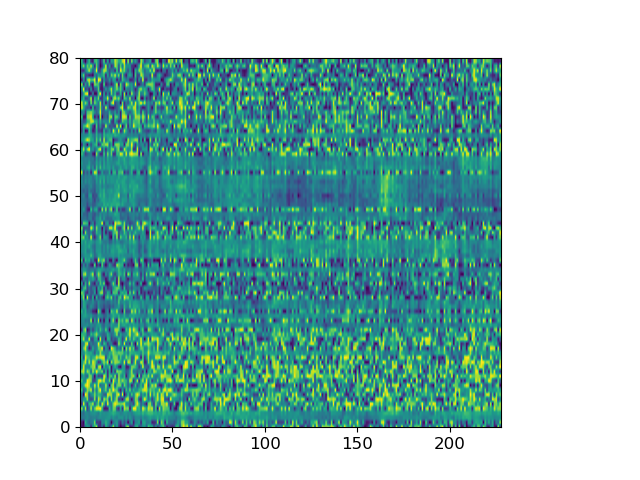}
      \label{tobe_single_band_tobe5000}
      \end{minipage}%
      }%
      \subfigure[In the 70th epoch.]{
        \begin{minipage}[t]{0.33\linewidth}
        \centering
        \includegraphics[width=2.0in]{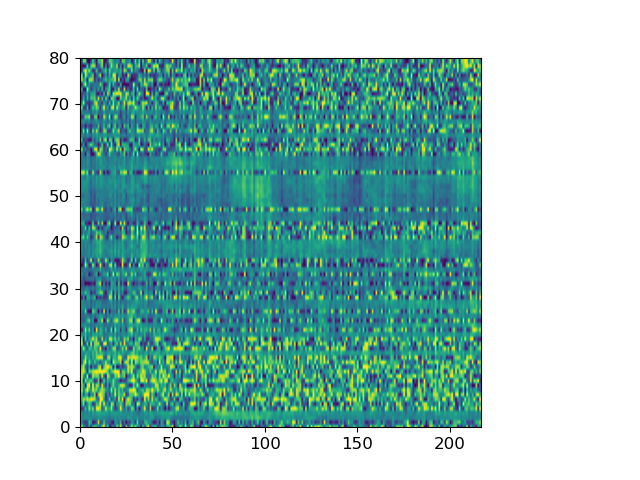}
        \label{tobe_single_band_tobe7000}
        \end{minipage}%
        }%
        \subfigure[In the 90th epoch.]{
          \begin{minipage}[t]{0.33\linewidth}
          \centering
          \includegraphics[width=2.0in]{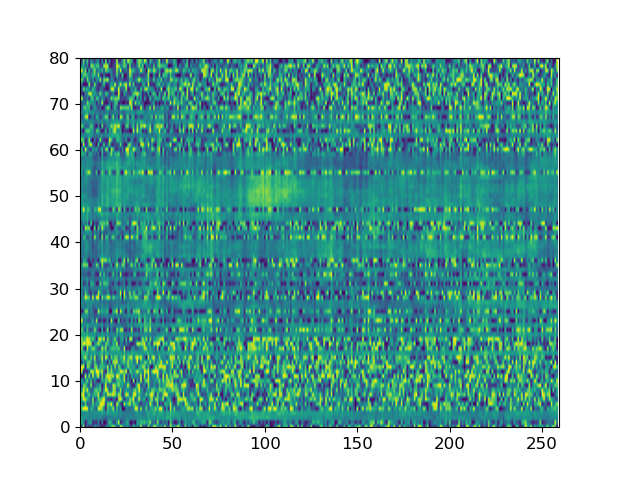}
          \label{tobe_single_band_tobe9000}
          \end{minipage}%
          }%

          \subfigure[In the 100th epoch.]{
            \begin{minipage}[t]{0.33\linewidth}
            \centering
            \includegraphics[width=2.0in]{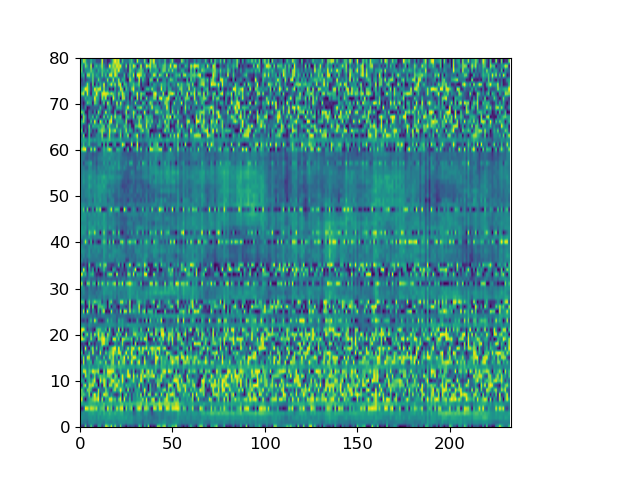}
            \label{tobe_single_band_tobe10000}
            \end{minipage}%
            }%
            \subfigure[In the 1000th epoch.]{
              \begin{minipage}[t]{0.33\linewidth}
              \centering
              \includegraphics[width=2.0in]{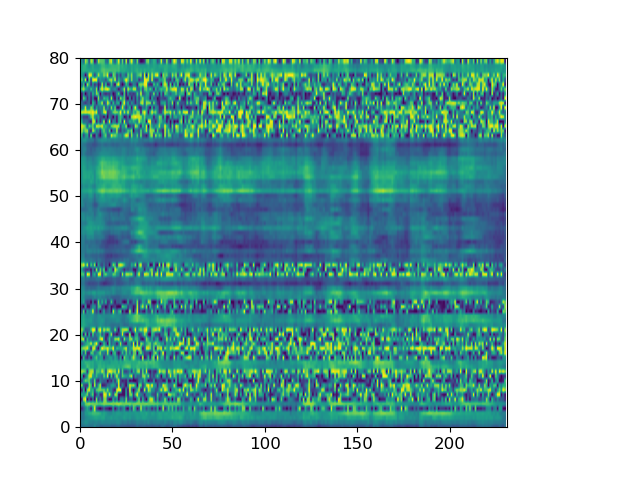}
              \label{tobe_single_band_tobe100000}
              \end{minipage}%
              }%
              \subfigure[In the 3000th epoch.]{
                \begin{minipage}[t]{0.33\linewidth}
                \centering
                \includegraphics[width=2.0in]{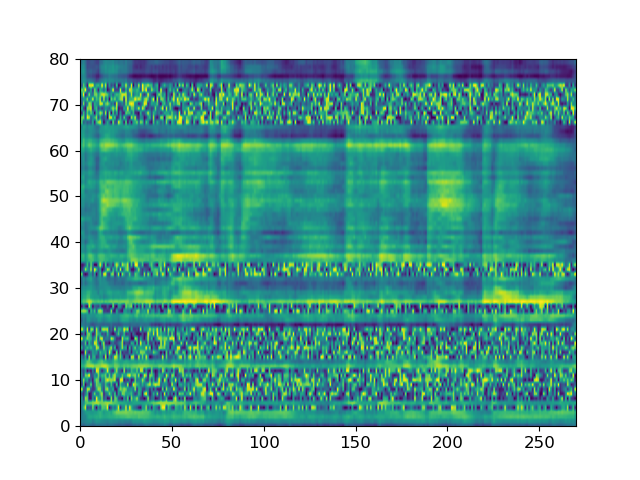}
                \label{tobe_single_band_tobe300000}
                \end{minipage}%
                }%

                \subfigure[In the 5000th epoch.]{
                  \begin{minipage}[t]{0.5\linewidth}
                  \centering
                  \includegraphics[width=2.0in]{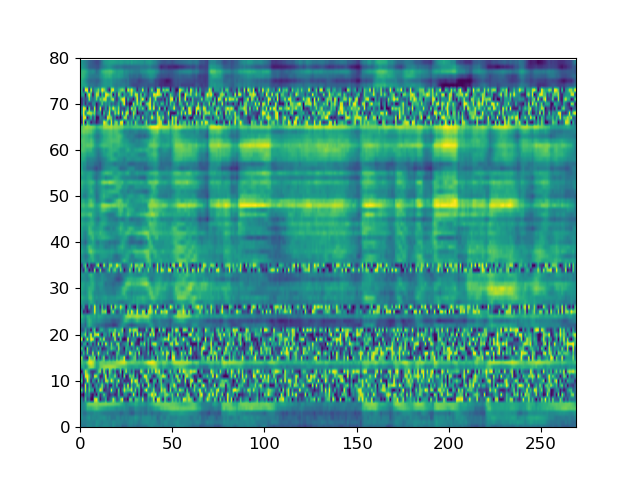}
                  \label{tobe_single_band_tobe500000}
                  \end{minipage}%
                  }%

  \centering
  \caption{In  It\^oTTS, only one decoder is used to decode all 80 channels in the
  mel spectrogram, the checkpoints obtained during the training process is used to generate the mel spectrogram, conditioned on the 
  text ``to be or not to be, this is a big problem''. }
 \label{tobe_single_band}
  \end{figure}

\begin{figure}[H]
    \centering
    \subfigure[Use the checkpoint obtained in the tenth epoch.]{
    \begin{minipage}[t]{0.33\linewidth}
    \centering
    \includegraphics[width=2.0in]{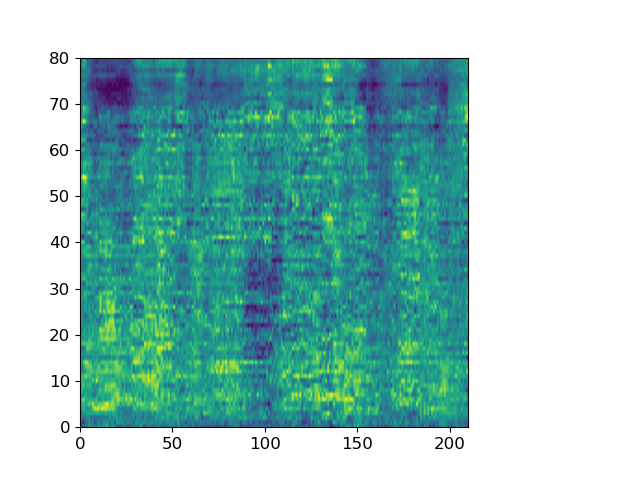}
    \label{tobe_single_band_tobe1000}
    \end{minipage}%
    }%
    \subfigure[In the 20th epoch.]{
    \begin{minipage}[t]{0.33\linewidth}
    \centering
    \includegraphics[width=2.0in]{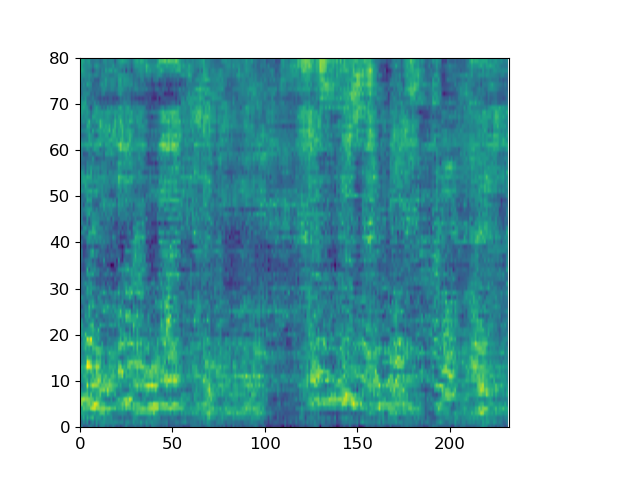}
    \label{tobe_single_band_tobe2000}
    \end{minipage}%
    }%
    \subfigure[In the 30th epoch.]{
      \begin{minipage}[t]{0.33\linewidth}
      \centering
      \includegraphics[width=2.0in]{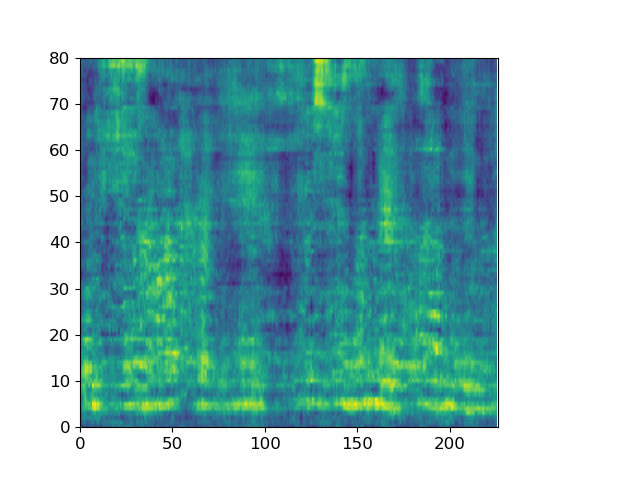}
      \label{tobe_single_band_tobe3000}
      \end{minipage}%
      }%
  
      \subfigure[In the 50th epoch.]{
        \begin{minipage}[t]{0.33\linewidth}
        \centering
        \includegraphics[width=2.0in]{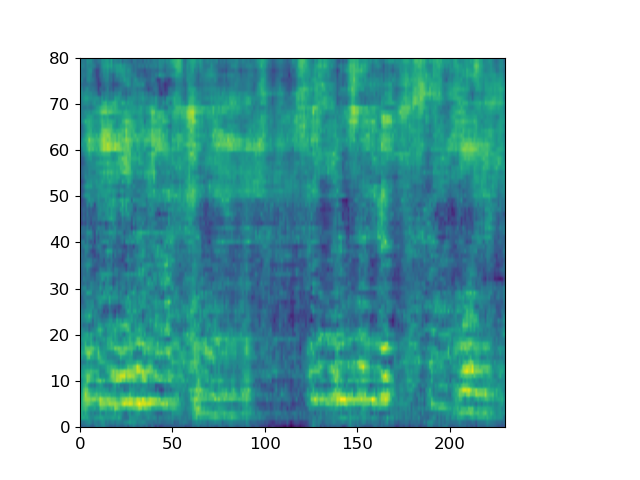}
        \label{tobe_single_band_tobe5000}
        \end{minipage}%
        }%
        \subfigure[In the 70th epoch.]{
          \begin{minipage}[t]{0.33\linewidth}
          \centering
          \includegraphics[width=2.0in]{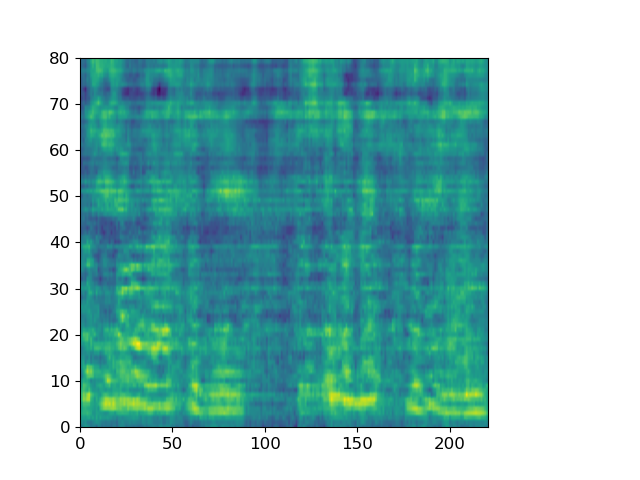}
          \label{tobe_single_band_tobe7000}
          \end{minipage}%
          }%
          \subfigure[In the 90th epoch.]{
            \begin{minipage}[t]{0.33\linewidth}
            \centering
            \includegraphics[width=2.0in]{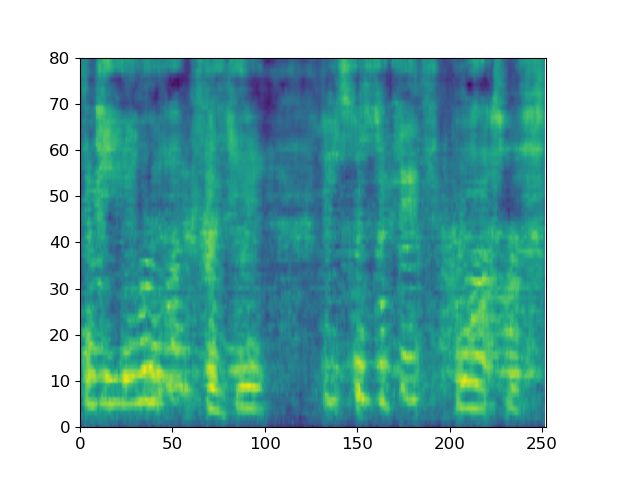}
            \label{tobe_single_band_tobe9000}
            \end{minipage}%
            }%
  
            \subfigure[In the 100th epoch.]{
              \begin{minipage}[t]{0.33\linewidth}
              \centering
              \includegraphics[width=2.0in]{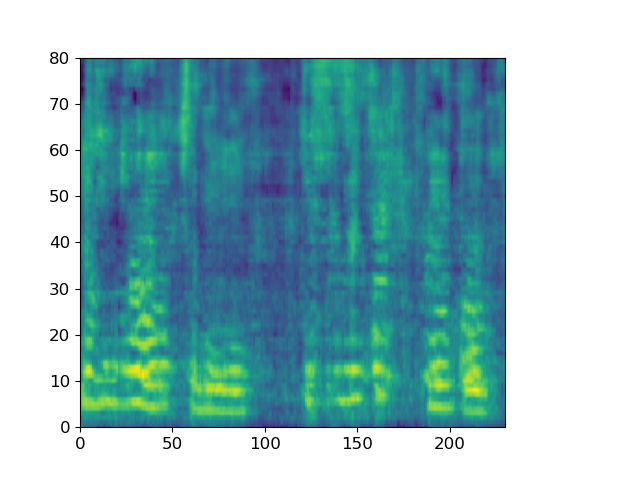}
              \label{tobe_single_band_tobe10000}
              \end{minipage}%
              }%
              \subfigure[In the 1000th epoch.]{
                \begin{minipage}[t]{0.33\linewidth}
                \centering
                \includegraphics[width=2.0in]{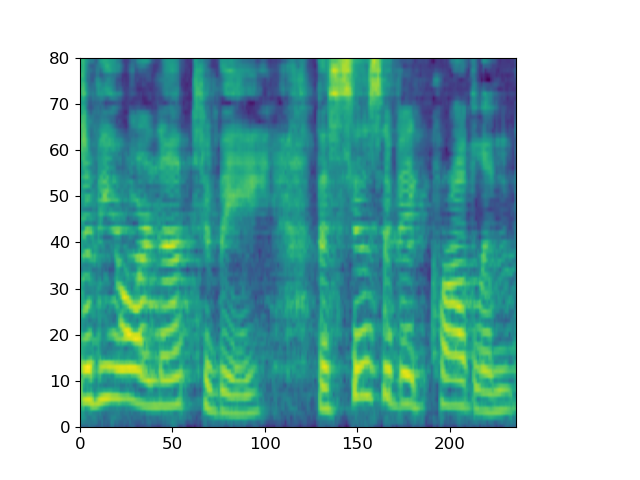}
                \label{tobe_single_band_tobe100000}
                \end{minipage}%
                }%
                \subfigure[In the 3000th epoch.]{
                  \begin{minipage}[t]{0.33\linewidth}
                  \centering
                  \includegraphics[width=2.0in]{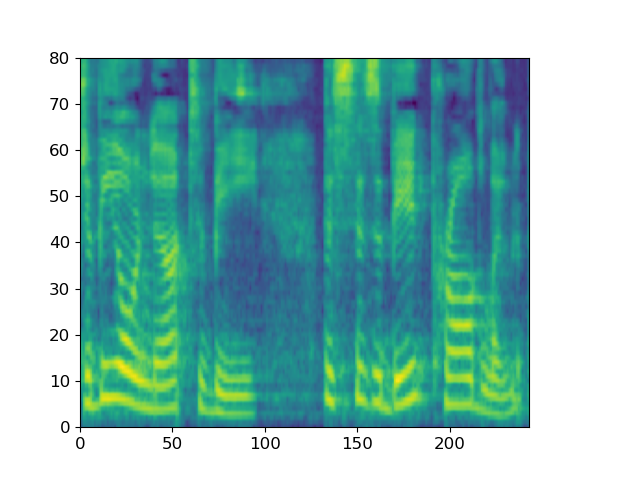}
                  \label{tobe_single_band_tobe300000}
                  \end{minipage}%
                  }%
  
                  \subfigure[In the 5000th epoch.]{
                    \begin{minipage}[t]{0.5\linewidth}
                    \centering
                    \includegraphics[width=2.0in]{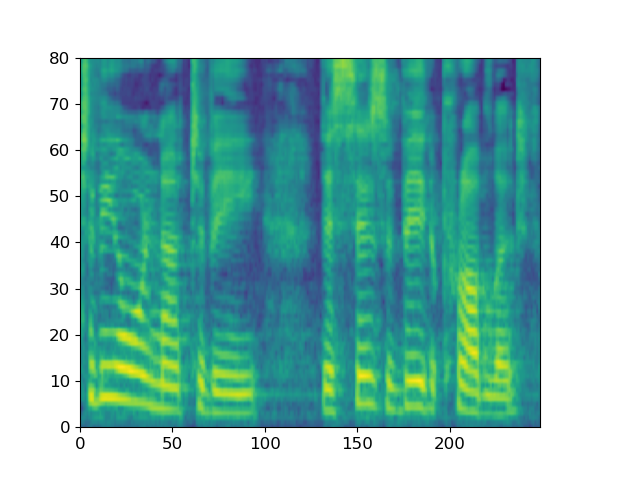}
                    \label{tobe_single_band_tobe500000}
                    \end{minipage}%
                    }%
  
    \centering
    \caption{
      In  It\^oTTS, several decoders are used, each decoder is only responsible for decoding the limited channels in the mel spectrogram. 
      For example, 8 decoders are used, 
      and each is only responsible for decoding 10 channels, and the checkpoint obtained during the training process is used to generate the mel spectrogram, 
      conditioned on the 
    text ``to be or not to be, this is a big problem''. }
   \label{tobe_multi_bands}
    \end{figure}

As shown in the Figure~\ref{tobe_single_band} and Figure~\ref{tobe_multi_bands}, if only one decoder is used for all the bands or channels of the mel spectrogram, 
the result is that the training cannot be achieved, and an interesting phenomenon 
is also found, that is, the network will first learn the easy-to-learn parts, such as the part without complex  harmonics textures.

\subsection{Diffusion generation of the mel spectrogram}

\begin{figure}[H]
  \centering
  \subfigure[Mel spectrogram in the first step.]{
  \begin{minipage}[t]{0.33\linewidth}
  \centering
  \includegraphics[width=2.0in]{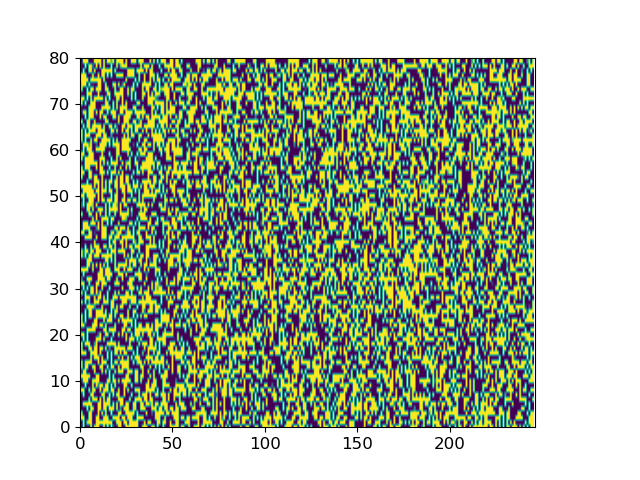}
  \label{tobe_itotts_diffusion_step1}
  \end{minipage}%
  }%
  \subfigure[At the 200th step.]{
  \begin{minipage}[t]{0.33\linewidth}
  \centering
  \includegraphics[width=2.0in]{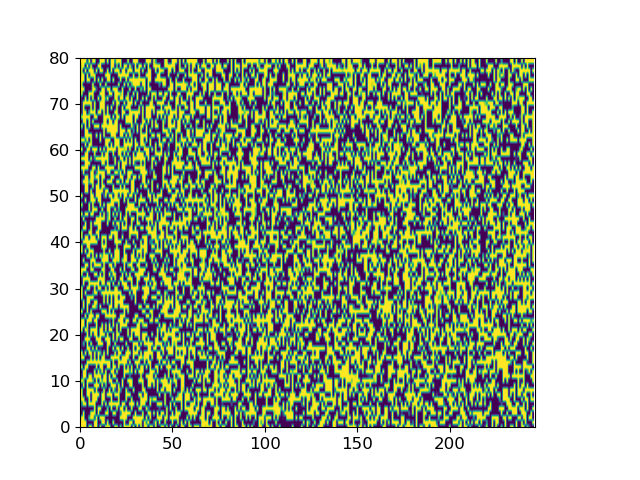}
  \label{tobe_itotts_diffusion_step200}
  \end{minipage}%
  }%
  \subfigure[At the 300th step.]{
    \begin{minipage}[t]{0.33\linewidth}
    \centering
    \includegraphics[width=2.0in]{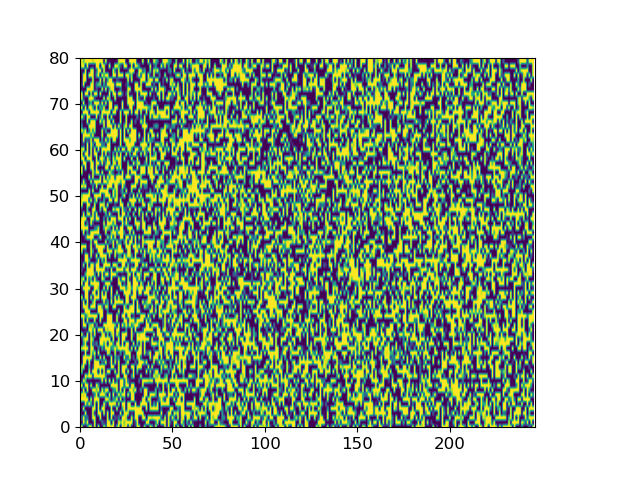}
    \label{tobe_itotts_diffusion_step300}
    \end{minipage}%
    }%

    \subfigure[At the 400th step.]{
      \begin{minipage}[t]{0.33\linewidth}
      \centering
      \includegraphics[width=2.0in]{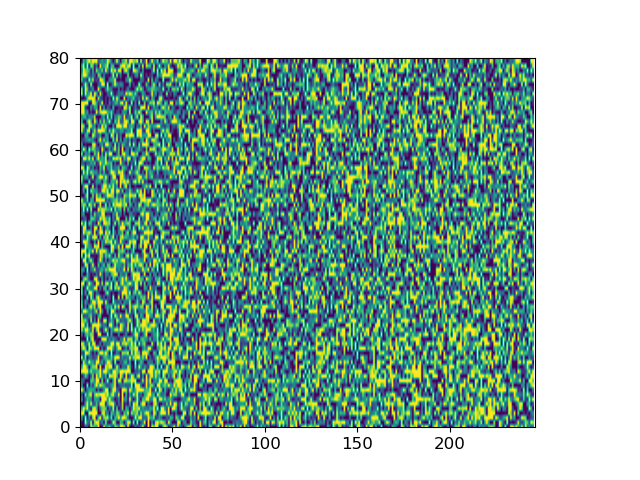}
      \label{tobe_itotts_diffusion_step400}
      \end{minipage}%
      }%
      \subfigure[At the 500th step.]{
        \begin{minipage}[t]{0.33\linewidth}
        \centering
        \includegraphics[width=2.0in]{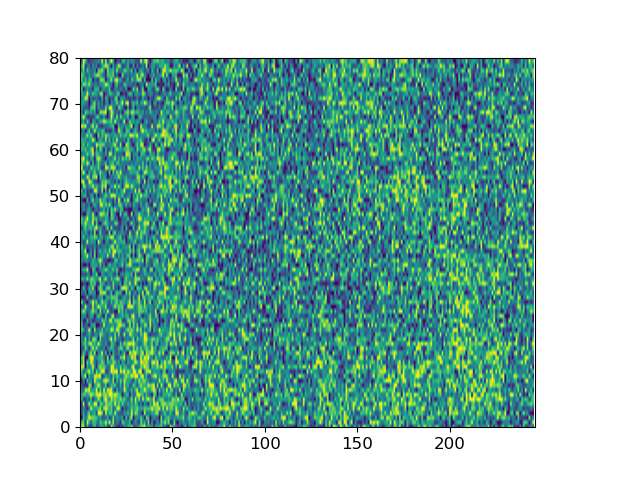}
        \label{tobe_itotts_diffusion_step500}
        \end{minipage}%
        }%
        \subfigure[At the 600th step.]{
          \begin{minipage}[t]{0.33\linewidth}
          \centering
          \includegraphics[width=2.0in]{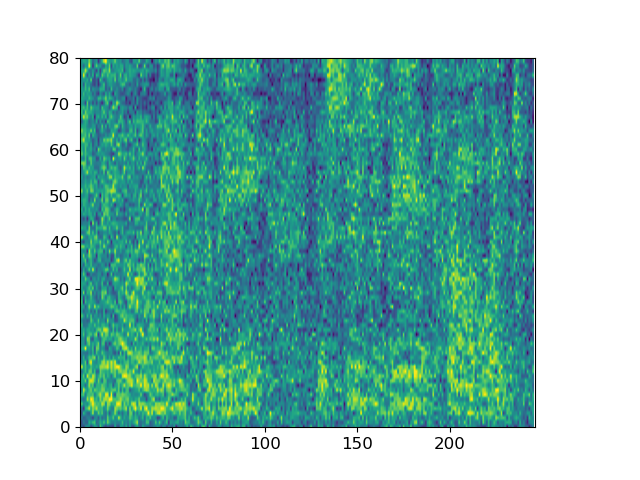}
          \label{tobe_itotts_diffusion_step600}
          \end{minipage}%
          }%

          \subfigure[At the 700th step.]{
            \begin{minipage}[t]{0.33\linewidth}
            \centering
            \includegraphics[width=2.0in]{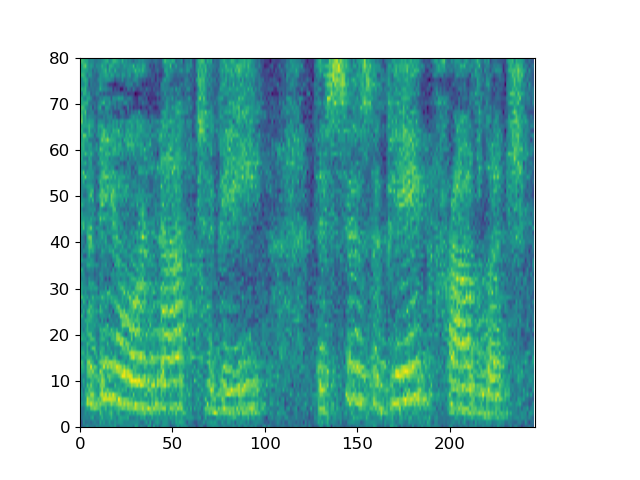}
            \label{tobe_itotts_diffusion_step700}
            \end{minipage}%
            }%
            \subfigure[At the 800th step.]{
              \begin{minipage}[t]{0.33\linewidth}
              \centering
              \includegraphics[width=2.0in]{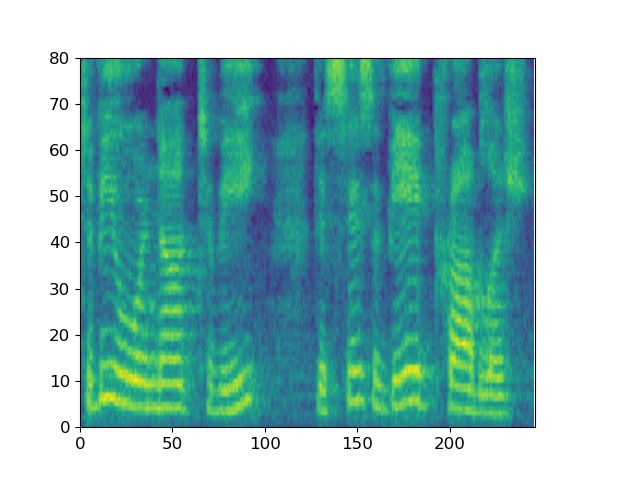}
              \label{tobe_itotts_diffusion_step800}
              \end{minipage}%
              }%
              \subfigure[At the 900th step.]{
                \begin{minipage}[t]{0.33\linewidth}
                \centering
                \includegraphics[width=2.0in]{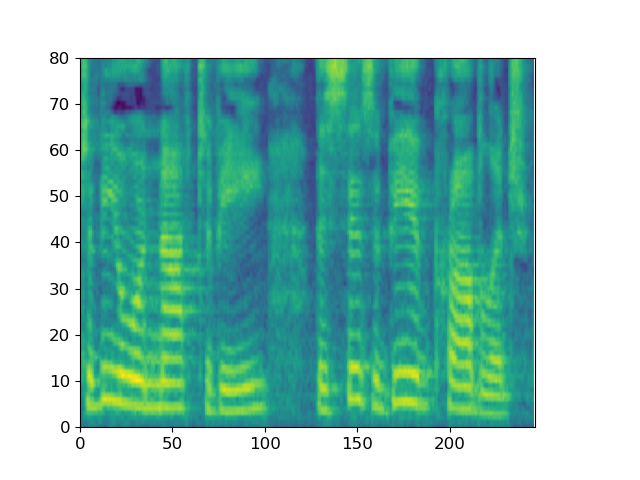}
                \label{tobe_itotts_diffusion_step900}
                \end{minipage}%
                }%

                \subfigure[Mel spectrogram at the 1000th step.]{
                  \begin{minipage}[t]{0.5\linewidth}
                  \centering
                  \includegraphics[width=2.0in]{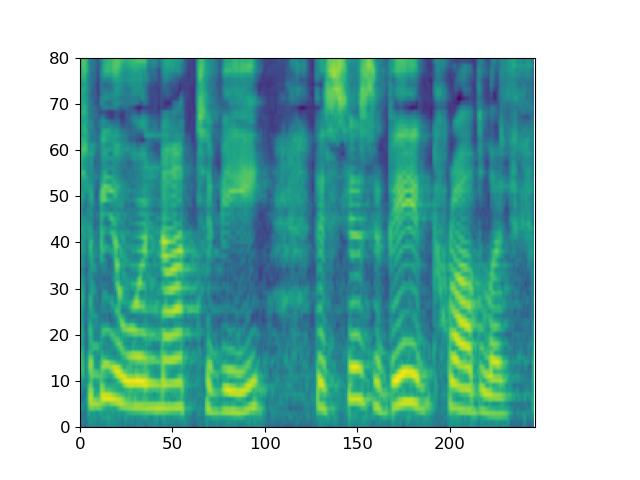}
                  \label{tobe_itotts_diffusion_step1000}
                  \end{minipage}%
                  }%

  \centering
  \caption{Conditioned on the text ``to be or not to be, this is a big problem'', It\^oTTS gradually generates the corresponding 
  mel spectrogram from the Gaussian signal step by step.}
 \label{tobe_itotts_diffusion}
  \end{figure}

As shown in the Figure~\ref{tobe_itotts_diffusion}, we can see how It\^oTTS gradually turns white noise into mel spectrogram.

\subsection{Diffusion generation of the wave}

\begin{figure}[th]
  \centering
  \subfigure[The waveform in the first step.]{
  \begin{minipage}[t]{0.5\linewidth}
  \centering
  \includegraphics[width=2.8in]{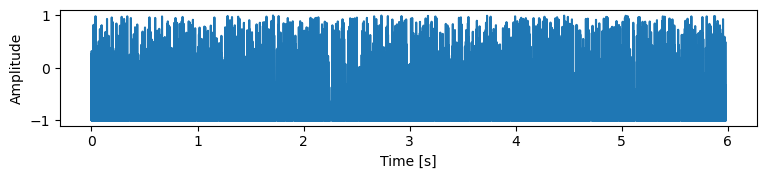}
  \label{tobe_itowave_diffusion_step1}
  \end{minipage}%
  }%
  \subfigure[The waveform at the 200th step.]{
  \begin{minipage}[t]{0.5\linewidth}
  \centering
  \includegraphics[width=2.8in]{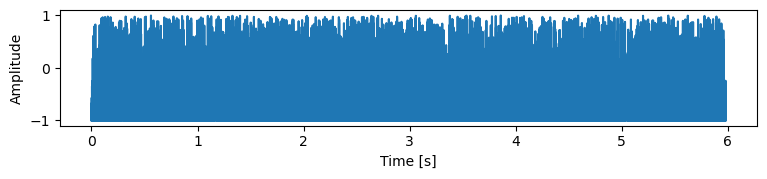}
  \label{tobe_itowave_diffusion_step200}
  \end{minipage}%
  }%

  \subfigure[The waveform at the 300th step.]{
    \begin{minipage}[t]{0.5\linewidth}
    \centering
    \includegraphics[width=2.8in]{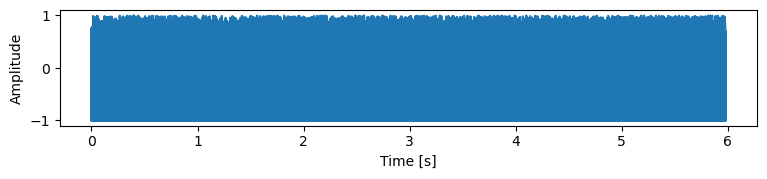}
    \label{tobe_itowave_diffusion_step300}
    \end{minipage}%
    }%
    \subfigure[The waveform at the 400th step.]{
      \begin{minipage}[t]{0.5\linewidth}
      \centering
      \includegraphics[width=2.8in]{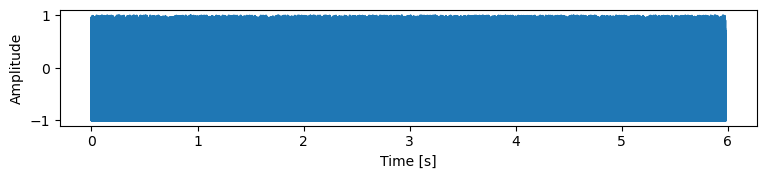}
      \label{tobe_itowave_diffusion_step400}
      \end{minipage}%
      }%

      \subfigure[The waveform at the 500th step.]{
        \begin{minipage}[t]{0.5\linewidth}
        \centering
        \includegraphics[width=2.8in]{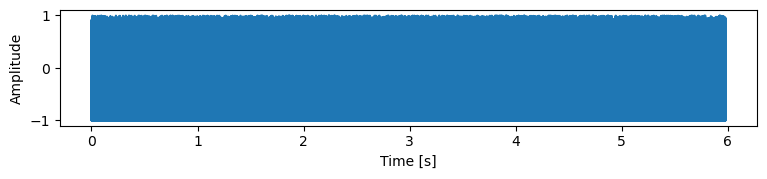}
        \label{tobe_itowave_diffusion_step500}
        \end{minipage}%
        }%
        \subfigure[The waveform at the 600th step.]{
          \begin{minipage}[t]{0.5\linewidth}
          \centering
          \includegraphics[width=2.8in]{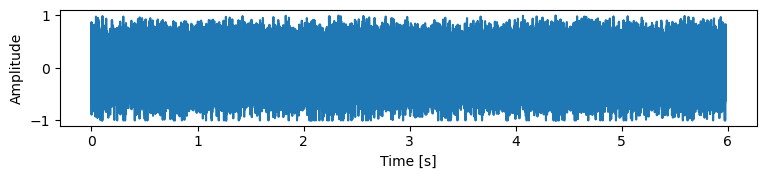}
          \label{tobe_itowave_diffusion_step600}
          \end{minipage}%
          }%

          \subfigure[The waveform at the 700th step.]{
            \begin{minipage}[t]{0.5\linewidth}
            \centering
            \includegraphics[width=2.8in]{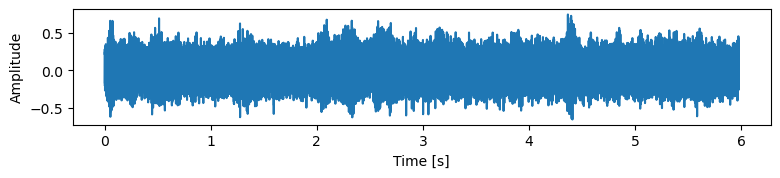}
            \label{tobe_itowave_diffusion_step700}
            \end{minipage}%
            }%
            \subfigure[The waveform at the 800th step.]{
              \begin{minipage}[t]{0.5\linewidth}
              \centering
              \includegraphics[width=2.8in]{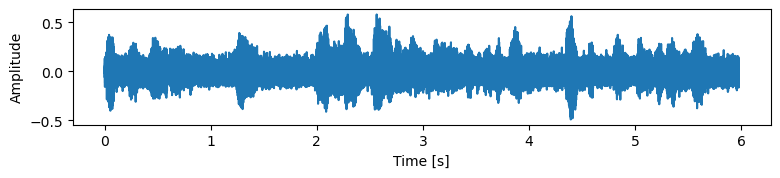}
              \label{tobe_itowave_diffusion_step800}
              \end{minipage}%
              }%

              \subfigure[The waveform at the 900th step.]{
                \begin{minipage}[t]{0.5\linewidth}
                \centering
                \includegraphics[width=2.8in]{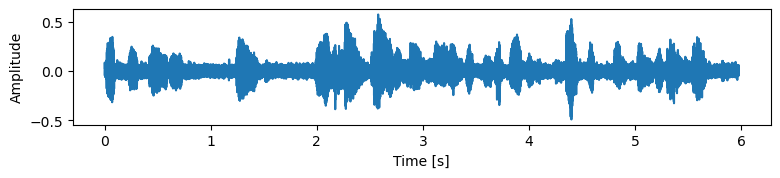}
                \label{tobe_itowave_diffusion_step900}
                \end{minipage}%
                }%
                \subfigure[The waveform at the 1000th step.]{
                  \begin{minipage}[t]{0.5\linewidth}
                  \centering
                  \includegraphics[width=2.8in]{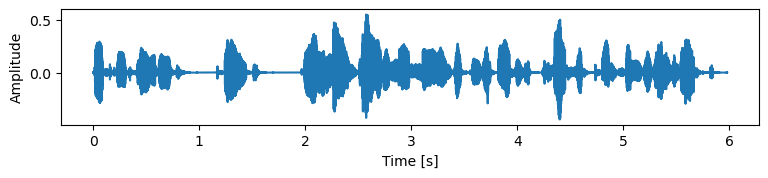}
                  \label{tobe_itowave_diffusion_step1000}
                  \end{minipage}%
                  }%

  \centering
  \caption{Conditioned on the frequency spectrum of the sentence LJ032-0167 in LJSpeech,  It\^oWave generates the corresponding voice step 
  by step from the Gaussian signal. The corresponding text is ``he concluded, quote, there is no doubt in my mind that these fibers 
  could have come from this shirt.''}
 \label{tobe_itowave_diffusion}
  \end{figure}

As shown in the Figure~\ref{tobe_itowave_diffusion}, we can see how It\^oWave gradually turns white noise into meaningful wave.

\end{document}